\documentstyle[multicol,aps,prl]{revtex}
\begin{document}
\widetext

\draft

\title{Elastic turbulence in curvilinear flows of polymer solutions.}
\author{Alexander Groisman$^a$ and Victor Steinberg$^b$}
\address{$^a$Department of Physics, UCSD, 9500 Gilman Dr., La
Jolla, CA 92093-0374,USA,\\
 $^b$Department of Physics of Complex
Systems, Weizmann Institute of Science, Rehovot 76100, Israel}




\date{\today}
\maketitle
\begin{abstract}
Following our first report (A. Groisman and V. Steinberg, $\sl
Nature$ $\bf 405$, 53 (2000)) we present an extended account of
experimental observations of elasticity induced turbulence in
three different systems: a swirling flow between two plates, a
Couette-Taylor (CT) flow between two cylinders, and a flow in a
curvilinear channel (Dean flow).  All three set-ups had high ratio
of width of the region available for flow to radius of curvature
of the streamlines. The experiments were carried out with dilute
solutions of high molecular weight polyacrylamide in concentrated
sugar syrups. High polymer relaxation time and solution viscosity
ensured prevalence of non-linear elastic effects over inertial
non-linearity, and development of purely elastic instabilities at
low Reynolds number (Re) in all three flows. Above the elastic
instability threshold, flows in all three systems exhibit features
of developed turbulence. Those include: (i)randomly fluctuating
fluid motion excited in a broad range of spatial and temporal
scales; (ii) significant increase in the rates of momentum and
mass transfer (compared to those expected for a steady flow with a
smooth velocity profile). Phenomenology, driving mechanisms, and
parameter dependence of the elastic turbulence are compared with
those of the conventional high Re hydrodynamic turbulence in
Newtonian fluids. Some similarities as well as multiple principal
differences are found. In two out of three systems (swirling flow
between two plates and flow in the curvilinear channel) power
spectra of velocity fluctuations are decaying rather quickly,
following power laws with exponents of about -3.5. It suggests
that being random in time, the flow is rather smooth in space, in
the sense that the main contribution to deformation and mixing
(and, possibly, elastic energy) is coming from flow at the largest
scale of the system. This situation - random in time and smooth in
space - is analogous to flows at small scales (below Kolmogorov
dissipation scale) in high Re turbulence.

\end{abstract}


\begin{multicols}{2}
\narrowtext

\section{ Introduction.}

Turbulence is one of the most fascinating phenomena in nature and one of the biggest challenges for
modern physics. It is common knowledge that a flow of a simple, low molecular, Newtonian fluid is
likely to be turbulent, when velocity, $V$, is high, kinematic viscosity of the fluid, $\nu$, is low,
and size of the tank, $L$, is large. All these three factors contribute to high value of the Reynolds
number, $Re=VL/\nu $. Motion of the Newtonian fluids is governed by the Navier-Stokes equation
\cite{landau}
\begin{equation}
\frac{\partial \vec{V}}{\partial t}+(\vec{V}\vec{\nabla})\vec{V}=-\nabla p/\rho +\nu \Delta \vec{V},
\end{equation}
where $p$ is pressure and $\rho $ is density of the fluid. The
equation has a non-linear term, $(\vec{V}\vec{\nabla})\vec{V}$,
which is inertial in its nature. The Reynolds number defines the
ratio of this
non-linear term to the viscous dissipative term, $\nu \Delta \vec{V}$. When $%
Re$ is high, the non-linear effects are strong, and the flow is likely to be turbulent. Therefore,
turbulence in fluids at high $Re$ is a paradigm for a strongly non-linear phenomenon in spatially
extended systems\cite{landau,Tritt}.

Solutions of flexible high molecular weight polymers are
visco-elastic liquids, and they differ from Newtonian fluids in
many aspects \cite{bird}. The most striking elastic property of
the polymer solutions is, probably, dependence of mechanical
stresses in flow on history of the flow. So, the stresses do not
immediately become zero when fluid motion stops, but rather decay
with some characteristic relaxation time, $\lambda $, which can be
well above a second. When a polymer solution is sufficiently
dilute, its stress tensor, $\mathbf{\tau }$, can be divided into
two parts, $\mathbf{\tau }=\mathbf{\tau }_{s}+\mathbf{\tau }_{p}$.
The first part, $\mathbf{\tau }_{s}$, is defined by viscosity of
the Newtonian solvent, $\eta _{s}$, and rate of strain in the
flow, $\mathbf{\tau }_{s}=-\eta
_{s}[\vec{\nabla}\vec{V}+(\vec{\nabla}\vec{V})^{T}]$. So, the
equation of motion for a dilute polymer solution becomes
\begin{equation}
\frac{\partial \vec{V}}{\partial t}+(\vec{V}\vec{\nabla})\vec{V}=-\vec{\nabla%
}p/\rho +(\eta_s/\rho) \Delta \vec{V}-\mathbf{\nabla \tau }{_{p}}/\rho
\end{equation}
Here the elastic stress tensor, $\mathbf{\tau }_{p}$, is due to
the polymer molecules, which are stretched in the flow, and it
depends on history of the flow. One can see that $\mathbf{\tau
}_{p}$ enters the equation of motion linearly. So the degree of
non-linearity of the equation of motion can still be defined by
the Reynolds number, $Re=VL\rho/\eta _{s}$.

The simplest model incorporating the elastic nature of the polymer
stress tensor, $\mathbf{\tau}_{p}$, is a Maxwell type constitutive
equation \cite {bird} with a single relaxation time, $\lambda $,
\begin{equation}
\mathbf{\tau }_{p}+\lambda {\frac{D\mathbf{\tau }_{p}}{Dt}}=-\eta _{p}[\vec{%
\nabla}\vec{V}+(\vec{\nabla}\vec{V})^{T}].
\end{equation}
Here ${\frac{D\mathbf{\tau }_{p}}{Dt}}$ is a material time
derivative of the polymer stress. An appropriate expression for
the time derivative $\frac{D\mathbf{\tau }_{p}}{Dt}$ has to take
into account that the stress is carried by fluid elements, which
move, rotate and deform in the flow. The translational motion
implies an advection term $(\vec{V}\vec{\nabla})\cdot \mathbf{\tau
}_{p}$ in an appropriate expression for
${\frac{D\mathbf{\tau}_{p}}{Dt}}$, while the rotation and
deformation of the fluid particles should lead to contributions
like $(\mathbf\vec{\nabla}\vec{V})\cdot {\tau }_{p}$ or
$\mathbf{\tau}_{p}\cdot (\vec{\nabla}\vec{V})$ \cite{bird}.
Therefore, along with terms linear in $\mathbf{\tau} _{p}$ and
$\mathbf\vec{V}$, some non-linear terms, in which
$\mathbf{\tau}_{p}$ is coupled to $\mathbf\vec{V}$,
should appear in the constitutive relation. A simple model equation for ${%
\frac{D\mathbf{\tau }_{p}}{Dt}}$, which is commonly used for
description of dilute polymer solutions, is the upper convected
time derivative,
\begin{equation}
{\frac{D\mathbf{\tau }_{p}}{Dt}}={\frac{\partial \mathbf{\tau }_{p}}{%
\partial t}}+(\vec{V}\vec{\nabla})\cdot \mathbf{\tau }_{p}-(\vec{\nabla}\vec{%
V})^{T}\cdot \mathbf{\tau }_{p}-\mathbf{\tau }_{p}\cdot (\vec{\nabla}\vec{V}%
).
\end{equation}
The equations (3,4) together with the expression for
$\mathbf{\tau}_{s}$ constitute the Oldroyd-B model of polymer
solution rheology \cite{bird}. One can see that non-linear terms
in the constitutive equation (Eqs. 3,4) are all of the order
$\lambda (V/L)\mathbf{\tau _{p}}$. The ratio of those non-linear
terms to the linear relaxation term, $\mathbf{\tau }_{p}$, is
given by a dimensionless expression $\lambda (V/L)$, which is
usually called Weissenberg number, $Wi$. (The relaxation term
$\mathbf{\tau }_{p}$ is somewhat analogous to the dissipation term
in the Navier-Stokes equation.)

One can expect mechanical properties of the polymer solutions to
become notably non-linear at sufficiently large Weissenberg
numbers. Indeed, quite a few effects originating from the
non-linear polymer stresses have been known for a long time
\cite{bird}. So, in a simple shear flow of a polymer solution
there is a difference between normal stresses along the direction
of the flow and along the direction of velocity gradient. At low
shear rates this normal stress difference, $N_{1}$, is
proportional to the shear rate squared. When flow lines are
curvilinear, it gives rise to a volume force acting on the liquid
in the direction of the curvature, the ''hoop stress''. Therefore,
if a rotating rod is inserted in an open vessel with a highly
elastic polymer solution, the liquid starts to climb up on the
rod, instead of being pushed outwards \cite{Weiss}. This
phenomenon is known as ''rod climbing'', or ''Weissenberg
effect''. Further, in an purely extensional flow resistance of a
polymer solution depends on rate of extension in a strongly
non-linear fashion. There is a sharp growth in the elastic
stresses, when the rate of extension exceeds $1/(2\lambda )$, that is at $%
Wi>1/2$. As a result apparent viscosity of a dilute polymer
solution can increase by up to three orders of magnitude
\cite{sridhar}. Both the Weissenberg effect and the growth of the
extensional flow resistance have been most clearly observed in
very viscous polymer solutions and in flows with quite low $Re$,
when non-linear inertial effects were insignificant.

A natural question arising here is, whether there may exist some
kind of turbulent flow produced by the non-linear polymer stresses
alone, in the absence of any significant inertial effects, at low
$Re$. An important step in this direction was made about a decade
and a half ago when purely elastic instabilities were
experimentally identified in a rotational flow between two plates
\cite{magda} and in the Couette-Taylor (CT) flow between two
cylinders \cite{LMS}. The experiments were carried out with a
Boger fluid \cite{boger}, a dilute solution of high molecular
weight polymers in a viscous Newtonian solvent. The Boger fluids
are almost universally used as model viscoelastic fluids. Their
relaxation times can be quite large, reaching seconds or even
minutes, while their rheological properties are
semi-quantitatively described by the simple Oldroyd-B model. The
purely
elastic instabilities occurred at $Wi$ of order unity and vanishingly small $%
Re$. As a result of the instabilities, secondary vortex flows
developed \cite {LMS}, and flow resistance somewhat increased
\cite{magda}. The problem of linear stability of the CT flow was
also treated theoretically using the Oldroyd-B model\cite{LMS2}.
The analysis showed that the non-linear mechanical properties of
the polymer solution can indeed lead to a flow instability, and a
simple mechanism of this purely elastic instability was proposed.

During the past decade the purely elastic instabilities in the
Boger fluids have been a subject of many theoretical and
experimental studies, which are partially reviewed in
Ref.\cite{Larson,Shaq}. After the pioneering work by Larson,
Muller and Shaqfeh \cite{LMS,LMS2}, purely elastic instabilities
were also found in other shear flows with curvilinear streamlines.
Those included the flow between a rotating cone and a plate and
the Taylor-Dean flow\cite{Larson,Shaq}. The original theoretical
analysis of Ref. \cite{LMS2} was refined, and more elaborate
experiments were carried out. A few new mechanisms of flow
instability driven by non-linear elastic stresses were suggested
for cone-and-plate and for Taylor-Dean flows. The original
mechanism proposed in Ref. \cite{LMS} was verified experimentally
in Ref.\cite{Ours2}.

Some flow patterns observed above the purely elastic instability
threshold in the curvilinear flows had rather disordered
appearance and exhibited chaotic spectra of velocities. So, it was
reasonable to suggest that at some conditions a truly turbulent
flow might be excited by elastic stresses at vanishingly small
$Re$. This idea was explicitly stated in Ref.\cite{EPL}, where
analogy between elastic and inertial flow transitions was
discussed. In fact, irregular flow patterns and growth of flow
resistance in elastic polymer solutions at low $Re$ were observed
even before the purely elastic instabilities were identified
\cite{Gies}. Those flow phenomena were even sometimes coined
''elastic turbulence''. In all those cases, however, the term
''turbulence'' was used in a rather loose fashion, without being
given a proper definition. More important, no quantitative data on
either flow velocity field or spatial and temporal velocity
spectra in those irregular flows were ever presented.

Although the notion of turbulence is widely used in scientific and
technical literature, there is no unique commonly accepted
definition of it. Therefore, turbulent flow is usually identified
by its main features. Turbulence implies fluid motion in a broad
range of temporal and spatial scales, so that many degrees of
freedom are excited in the system. There are no characteristic
scales of time and space in the flow, except for those restricting
the excited temporal and spatial domains from above and below.
Turbulent flow is also usually accompanied by a significant
increase in momentum and mass transfer. That is, flow resistance
and rate of mixing in a turbulent flow become much higher than
they would be in an imaginary laminar flow with the same $Re$.

In our recent publication \cite{Nat} we showed, how the first
three of these features of turbulence appeared in a flow of a
highly elastic polymer solution at low Reynolds numbers. The
experiments were done in a swirling flow between two plates with a
wide gap, and the phenomenon was called elastic turbulence. In the
present manuscript we give a more complete account of our
experiments on the elasticity driven turbulent flow. We show
additional velocity spectra, and distributions of probabilities of
flow velocities and their derivatives. We also show that the
elastic turbulence leads to quite efficient mixing in the flow.
Further, we present results of our experiments on Couette-Taylor
(CT) flow and on an open flow in a curvilinear channel (Dean
flow), where we also observed the elasticity driven turbulence.

The paper is organized as follows. In the next section we discuss
some practical problems concerning the experiment and the polymer
solution. In Sections III, IV and V the results on the swirling
flow between two plates, the CT flow and the flow in the
curvilinear channel, respectively, are presented. The results are
briefly summarized and discussed in Section VI.

\section{ Experimental Considerations.}

As it was suggested in Ref. \cite{EPL}, where the CT flow was
discussed, there is some analogy between flow transitions driven
by elasticity and inertia. So, the inertially driven Taylor
instability occurs at constant Taylor number \cite{landau,Tritt},
$Ta={\frac{d}{R}}Re^{2}$, while the elastic instability is
controlled by a parameter $K={\frac{\eta _{p}}{\eta
}}{\frac{d}{R}}Wi^{2}$ \cite{LMS2,Ours2}. Here $R$ is radius of
the inner cylinder, $d$ is width of the gap between the cylinders
in the CT set-up, $\eta $ is the total solution viscosity, and
$\eta _{p}$ is polymer contribution to the viscosity, $\eta
_{p}\equiv \eta -\eta _{s}$. The Weissenberg number is defined
here as $Wi=\lambda \Omega R/d$, where $\Omega $ is angular
velocity of the rotating inner cylinder. (It was termed as the
Deborah number in some of the original texts
\cite{LMS2,Ours2,EPL}.) The Weissenberg number, which reflects
non-linearity of the constitutive equation (Eqs. 3,4), appears to
be analogous to the Reynolds number, which expresses non-linearity
of the equation of motion. The geometric parameter determining
curvature, the gap ratio, $d/R$, enters the expressions for both
$Ta$ and $K$. (There is also a specific term $\eta _{p}/\eta $,
though, which shows polymer contribution to the solution viscosity
and naturally appears only in the expression for the parameter $K$
for the elastic instability). Scales of time and velocity for the
purely elastic flow transition are given by $\lambda $ and
$d/\lambda $, respectively. As it was shown in Ref. \cite{EPL}
they are analogous to $t_{vd}$ and $d/t_{vd}$, which define
scales of time and velocity for the inertially driven flow
transitions. Here $t_{vd}$ is the viscous diffusion time defined
as $t_{vd}=d^{2}/\nu $.

Nevertheless, along with all those analogies,there are still some
important differences between flow transitions driven by inertia
and elasticity. So, it is an experimental fact that \textit{any}
laminar flow of a Newtonian fluid becomes unstable at sufficiently
high $Re$, and all high Reynolds number flows are turbulent. That
includes rectilinear shear flows, such as Poiseille flow in a
circular pipe and plane Couette flow, which are supposed to be
linearly stable at any $Re$. In contrast to it, purely elastic
flow instabilities in shear flows have only been observed so far
in systems with curvilinear stream lines. All these instabilities
are supposed to be driven by the hoop stress, which originates
from the normal stress differences. Elastic instabilities also
occur in flows of polymer solutions with open surfaces and in
extensional flows through contractions in channels
\cite{Larson,rheol}. A particular feature of the latter flows is
that the rate of strain changes along the flow lines, so that even
the basic flow states are not stationary in the Langrangian sense.
Instabilities in those flows are not discussed in this paper.

This difference between the inertial and elastic instabilities may
originate, of course, from the distinct governing equations. There
are, however, some purely practical factors that can explain
rather well the lack of observations of purely elastic flow
transitions in rectilinear shear flows. Inertial instabilities in
rectilinear shear flows of Newtonian fluids occur at quite high
Reynolds numbers. Those are typically about two orders of
magnitude higher than $Re$ at which curvilinear shear flows with
large gap ratios, $d/R$, become unstable. A priory, one may
suggest that rectilinear and curvilinear shear flows would have a
similar relation between $Wi$ at thresholds of the purely elastic
flow instabilities as well. The problem is, however, that while it
is rather easy to generate high $Re$ flows with low viscosity
Newtonian fluids, it is usually impossible to reach the
corresponding high values of $Wi$ in shear flows of elastic
polymer solutions. That is, there always exist rather severe
practical limitations restricting non-linearity in elastic polymer
stresses in shear flows. Their molecular mechanisms have been
recently elucidated in a seminal paper by Chu's group
\cite{Chu-shear}.

Polymer molecules have finite extensibility, and their relaxation
time decreases when they get stretched in a shear flow. This
thinning of the relaxation time at high $Wi$ is usually quite a
strong and well recognized effect. In fact, different variations
of the basic Oldroyd-B model have been specially developed to take
into account the shear thinning \cite{bird}. Significant decrease
of $\lambda $ with shear rate, $\dot{\gamma}$, renders growth of
the Weissenberg number, $Wi=\lambda (\dot{\gamma})\dot{\gamma}$,
much slower than linear in $\dot{\gamma}$ . Substantial stretching
of the polymer molecules in the primary shear flow also reduces
their ability for further extension and susceptibility of the
elastic stresses to flow perturbations, which is necessary for
generation of the flow instabilities and secondary vortex
flows\cite{LMS2}. Finally, high shear rates cause mechanical
degradation of the polymer molecules. It leads to permanent
reduction of elasticity during experimental runs and decay of
$\lambda $ that can be very fast at high $Wi$. Because of all
those reasons it was found to be very difficult or even impossible
to observe elastic instabilities, when expected values of $Wi$ at
the instability threshold were high. It was the case in
curvilinear flows with small gap ratios, $d/R$, \cite{McK} and
small viscosity ratios $\eta _{p}/\eta $ \cite{Ours2}.

Therefore, in order to maximize the non-linear elastic effects and
to get a better opportunity to observe the elastic turbulence we
had to choose experimental conditions quite carefully \cite{Nat}.
First, it was important to obtain an elastic instability at\ a
possibly low critical Weissenberg number, $Wi_{c}$. For that
purpose the gap ratio and the viscosity ratio had to be possibly
large. Therefore, we used a polymer solution with a rather large
$\eta _{p}/\eta $ of about 1/4. (Further increase of the polymer
concentration and of $\eta _{p}/\eta $ was not very efficient, and
would also complicate the solution rheology, including large shear
thinning of the solution viscosity.) In order to have a large gap
ratio with a possibility to vary it, we carried out our
experiments in a swirling flow between two plates. The ratios
$d/R$ which we used were 0.263 and 0.526. In fact, historically we
first tested a CT flow with $d/R=1/2$, but growth of flow
resistance in it was significantly smaller than in the swirling
flow set-up, which we studied later (see sections III-IV).

Second, an appropriate polymer sample for the solution had to be
chosen, that would not suffer major mechanical degradation under
experimental flow conditions. Further, the limit for extensibility
of the polymer molecules had to be possibly high compared with
their typical conformations at the instability threshold. We used
Polyacrylamide (PAAm) with a large average weight
$M_{w}=18,000,000$ and a broad molecular weight distribution (and
low concentration of polyacrylic acid monomers) supplied by
Polysciences. This commercial polymer sample proved to be
remarkably stable with respect to mechanical degradation, that
allowed us to reach high values of $Wi$ and to explore strongly
non-linear flow regimes. The high molecular weight of the polymers
resulted in a large characteristic relaxation time, $\lambda $,
even with solvents of moderate viscosity, and in a small
characteristic stress, $\tau _{0}=\eta _{s}/\lambda $. One can
learn from the Eqs. 3,4 that $\tau _{0}$ sets a scale of the
polymer stress, at which its non-linearity becomes significant.
Therefore, the value of $\tau _{0}$ together with $\eta _{p}/\eta
$ and $d/R$ determine the polymer stress $\tau _{p}$ in the
primary shear flow at the instability threshold. It is rather
natural to suggest that when polymer molecules transduce less
stress, they are also less subjected to mechanical degradation.
Further, molecular interpretation of the Weissenberg number in a
shear flow is the degree of deformation of polymer molecules from
their relaxed random coil conformations. So, $Wi=1$ can be
regarded as a characteristic value at which extension of polymer
molecules becomes considerably larger than the size of relaxed
coils. By the same token, extension of a polymer molecule at
$Wi_c$ (the elastic instability threshold) is supposed to be a
fixed multiple of the relaxed coil size (in the first, linear
approximation valid at moderate $Wi_c$\cite{Chu-shear}). High
molecular weight and flexibility of a polymer suggests large
number of the Kuhn segments \cite{bird,Doi} in the polymer chain,
and a high ratio between its contour length (size, when fully
extended) and the size (radius of gyration, $R_g$) of a relaxed
coil. (For a polymer molecule in a good solvent, with $R_g\sim
M_{w}^{3/5}$, this ratio should increase as $M_{w}^{2/5}$
\cite{Doi}. We would like to point out here that addition of 1\%
of NaCl to the solution significantly reduced $R_g$.) Therefore,
using a higher $M_{w}$\ polymer we increase "reserve" of
extensibility starting from the typical conformation at $Wi_c$.
That opens a way for reacher flow dynamics above the elastic instability threshold.\\

\section{ Swirling flow between two plates.}

\subsection{ Experimental set-up and procedure.}

The experimental apparatus is schematically shown in Fig.1. Polymer solution was held in a stationary
cylindrical cup with a flat bottom (lower plate). A coaxial rotating upper plate was just touching the
surface of the fluid. The cup was mounted on top of a commercial rheometer, AR-1000 of TA-instruments.
The upper plate was attached to the shaft of the rheometer, which allowed precise control of its
rotation velocity, $\Omega$, and measurements of the torque, $T$. The average shear stress at the
upper plate, $\tau_{w}$, was
calculated using the equation $T=\tau_{w}\int rdS$, that gave $%
\tau_{w}\equiv3T/(2\pi R^3)$. (The integration is over the upper
plate surface.)

Sidewalls of the cup were machined of a single piece of perspex,
which was optically clear. The cup was circular from the inside
and square from the outside in horizontal cross-section. That
allowed measurements of flow velocity in the horizontal plane by a
laser Doppler velocimeter (LDV) with two crossing frequency
shifted beams. By appropriate positioning and orientation of the
beam crossing region, azimuthal (longitudinal) and radial
(spanwise) velocity components, $V_{\phi }$ and $V_{r}$,
respectively, could be measured at different $r$ and $z$. Here
$(r,\phi ,z)$ are cylindrical coordinates. The bottom of the cup
was machined of stainless steel and the temperature was stabilized
at 12 $^{\circ }$C by circulation of water below the bottom plate.

A slightly modified version of the set-up was designed to
photograph the flow from below and to observe mixing in the flow.
A special cup of the same shape but with a transparent bottom
(lower plate, made of perspex) was attached to the rheometer
concentrically with the shaft but above the rheometer base, and a
mirror tilted by 45$^{\circ }$ was placed under the cup, as it is
schematically shown in Fig.1. The mirror was used both to
illuminate the fluid by diffuse light and to relay images of the
flow to a CCD camera. The camera was equipped with a regular video
lens and mounted horizontally near the rheometer (Fig.1). The
source of the diffuse light was an illuminated white screen around
the camera. The images were digitized by an 8 bit 512x512 frame
grabber. In order to provide thermal stabilization, in this case
the whole rheometer was placed in a thermally isolated box with
through flow of temperature controlled air.

In the basic standard set-up, the radii of the upper plate and of
the cup were $R=38$ mm and $R_{2}=43.6$ mm, respectively, and the
distance between the plates was $d=10$ mm. The configuration was
similar to the devices with rotating flow between two plates used
in experiments on purely elastic
instability\cite{magda,Shaq,McK,McKinley}. Its gap ratio, $d/R$,
was significantly higher, though. In order to study dependence of
flow conditions on size of the system, two smaller set-ups, a
half-size and a quarter-size, with all the dimensions reduced by
factors of two and four, respectively, were used. Every time, when
dimensions of an experimental set-up are non-standard, it is
stated explicitly in the text.

We prepared a stock solution of PAAm ($M_{w}=18,000,000$ by
Polysciences) and used it through the whole course of the
experiments. First we dissolved 0.9 g of PAAm powder and 3 g of
NaCl in 275 ml of deionized water by gentle shaking. (Addition of
NaCl was necessary to improve solubility of PAAm.) Next the
solution was mixed for 3 hours in a commercial mixer with a
propeller at a moderate speed. The idea behind that procedure was
to cause mechanical degradation PAAm molecules with the highest
weights, and to "cut" a high $M$ tail of the broad molecular
weight distribution of the PAAm sample. In a solution with a broad
distribution of polymer molecular weights the heaviest molecules,
which are most vulnerable to mechanical degradation, may also make
a major contribution to the solution elasticity. A possible
negative effect of that is significant degradation of elasticity
during experimental runs, and inconsistency of results of the
experiments. We found empirically that the procedure of
pre-degradation in the mixer leads to substantial reduction of the
degradation during the experiments and to substantial improvement
of their consistency\cite{master}. Finally, 9 g of iso-Propanol
was added to the solution (to preserve it from aging) and water
was added up to 300 g. The final concentrations of PAAm, NaCl and
iso-Propanol in the stock solution were 3000 ppm, 1\% and 3\%,
respectively.

As a viscous Newtonian solvent for PAAm, we used a solution of
65\% sugar (saccharose) and 1\% NaCl in water. The salt was added
to fix the ionic contents. Viscosity and relaxation time were
measured with the same AR-1000 rheometer in a temperature
controlled narrow gap Mooney-Ewart geometry. The solvent viscosity
was $\eta _{s}=0.324$ Pa$\cdot $s at 12 $^{\circ }$C. The PAAm
concentration was 80 ppm, Fig.2. Solution viscosity,
$\eta(\dot{\gamma})$, was carefully measured in a broad range of
shear rates, $\dot{\gamma}$, Fig.2. Viscosity, $\eta $, was slowly
decreasing with $\dot{\gamma}$, so that its shear thinning was
about 7\% per a decade of $\dot{\gamma}$. At a shear rate of $\dot{\gamma}=1$ s$%
^{-1}$, corresponding to the onset of a purely elastic instability in the standard configuration
 (see below), $\eta $ was $0.424$ Pa$\cdot $s, and the
viscosity ratio was $\eta _{p}/\eta =0.235$. Polymer relaxation
time, $\lambda $, was measured in oscillatory tests with a shear
rate amplitude of 1 s$^{-1}$. Components of the apparent viscosity
of the solution in phase and out of phase with the applied shear,
$\eta'$ and $\eta''$, respectively, were measured in long series
of oscillations in a range of angular frequencies, $\omega$.
(Individual series were typically repeated about 10 times, and
average viscosity values were calculated.) The same procedure was
applied then to the solvent, and its viscosity components,
$\eta'_s$ and $\eta''_s$, were calculated as well. (The latter was
nearly zero and it was an important test of virtual absence of
inertial effects and of general applicability of the procedure.)
The values for the polymer in phase and out of phase viscosity
were calculated then as, $\eta'_p=\eta'-\eta'_s$ and
$\eta''_p=\eta''-\eta''_s$. The (frequency dependent) relaxation
time was calculated as
$\lambda(\omega)=tan^{-1}(\eta''_p/\eta'_p)/\omega$, and
$\lambda(\omega)$ at $\omega\rightarrow 0$, estimated as 3.4 s,
was chosen as a representative relaxation time, $\lambda$, inset
in Fig.2. The diffusion coefficient for the saccharose molecules
was about $D=8.5\cdot 10^{-7}$ cm$^{2}$/s \cite{Critables}.

In a swirling flow between two plates shear rate is quite
inhomogeneous over the fluid bulk, even when the flow is laminar.
So, in the simplest case of a narrow gap, $d/R<<1$, when the shear
rate is constant along the $z$-direction, it grows along the
radius, from zero at $r=0$ to the maximal value of $\Omega R/d$ at
$r=R$. Shear rate averaged over the surface of the upper plate
(and over the fluid volume) is then $2\Omega R/(3d)$. In our case,
when the gap is quite wide ($d/R=0.263$ in the standard
configuration) and the sidewalls are close to the upper plate,\
the situation is even more complicated. The shear rate becomes
strongly non-homogeneous along $z$ in a large region of space
corresponding to $r$ beyond $R-d$. So, measurements of the ratio
$T/\Omega $ in a laminar flow of a viscous Newtonian liquid gave a
value 1.68 times larger, than a value calculated suggesting
homogeneous shear rate along $z$. (The latter suggestion is not
realistic for the large gap ratio set-up, of course.) This
discrepancy is obviously due to a high shear rate layer near the
upper plate at large $r$. One can see that choice of a
representative shear rate becomes somewhat arbitrary under these
circumstances. We decided to consider the simple expression
$\Omega R/d$ as a characteristic shear rate, and to define the
Weissenberg number as $Wi=\lambda \Omega R/d$. The Reynolds number
was defined as $Re=\Omega Rd\rho/\eta$.

In order to evaluate growth of flow resistance due to elastic
instability and irregular secondary flow in the system, the
average shear stress near the upper plate, $\tau _{w}^{lam}$, in
an imaginary laminar shear flow at the same $\Omega $ had to be
estimated. The stress, $\tau _{w}^{lam}$, would depend on an
average shear rate, $\dot{\gamma}_{av}$, at the upper plate, and
on viscosity of the polymer solution, $\eta $, at
this shear rate. An appropriate expression for it is $\tau _{w}^{lam}=\eta (%
\dot{\gamma}_{av})\dot{\gamma}_{av}$. The average shear rate, $\dot{\gamma}%
_{av}$, was estimated from measurements of the ratio $\tau
_{w}/\eta _{0}$ in laminar flow of a Newtonian fluid with a large
viscosity, $\eta _{0}$, at
low $Re$. The shear rate in the laminar flow, calculated as $\dot{\gamma}%
_{av}=\tau _{w}/\eta _{0}$, was proportional to $\Omega $ , being $\dot{\gamma}%
_{av}=1.12\Omega R/d$ in the standard configuration, and it gave a
properly weighted average over the surface of the upper plate.

A suitable parameter characterizing growth of the flow resistance is the
ratio of the average stress at the upper plate, $\tau _{w}$ (defined as $%
\tau _{w}\equiv 3T/(2\pi R^{3})$ again), to the stress $\tau
_{w}^{lam}$ in the laminar flow at the same $\Omega $. In
Newtonian fluids this ratio generally increases with $Re$ as the
flow becomes increasingly irregular, and magnitude of $\tau
_{w}/\tau _{w}^{lam}$ can be considered as a measure of strength
of turbulence and of the turbulent resistance. In our standard
set-up, when a low viscosity Newtonian fluid is used, $\tau _{w}$
becomes 30\% higher than $\tau _{w}^{lam}$ at $Re=70$. Therefore,
$Re=70$ can be regarded as a characteristic value, at which
inertial effects in the flow become considerable.

\subsection{ Observation of elastic turbulence.}

\subsubsection{ Flow resistance.}

The first indication of a strongly non-linear state in the swirling flow was significant growth of the
flow resistance above the elastic instability
threshold. Dependence of $\tau_w/\tau_w^{lam}$ on the shear rate, $\dot{%
\gamma}=\Omega R/d$, is shown in Fig. 3. The shear rate was
gradually increased in time, very slowly (by about 10\% h$^{-1}$)
in the transition region, $\dot{\gamma}=0.8-1.1$ s$^{-1}$, and
faster below and above it. One can see a sharp transition in flow
of the polymer solution (curve 1, black
line), that occurs at $\dot{\gamma}$ of about 1 s$^{-1}$ (corresponding to $%
Wi=3.5$), and that is manifested in a significant increase in
$\tau _{w}$ compared to the laminar flow. The Reynolds number at
the transition point is about 0.3, so that the inertial effects
are quite negligible there. The transition has
pronounced hysteresis (gray line, corresponding to slow reduction of $\dot{%
\gamma}$), which is rather typical for the purely elastic flow
instability \cite{Ours2}. The ratio $\tau_w/\tau_w^{lam}$ keeps
growing with the shear rate and at the highest $\dot{\gamma}$ that
has been reached the flow resistance is about 12 times larger than
it would be in a laminar flow. In the same range of shear rates
flow of the pure solvent, curve 3, is completely laminar and the
ratio $\tau_w/\tau_w^{lam}$ is unity within resolution of the
rheometer (about 1\%).

Growth of the resistance in the polymer solution flow becomes even
larger, when the size of the gap is increased to 20 mm (2nd curve
in Fig.3), and the gap ratio becomes 0.526. Then the ratio
$\tau_w/\tau_w^{lam}$ reaches a value of 19. For Newtonian fluids
in the same flow geometry such growth of the flow resistance is
found at $Re$ of about 2$\cdot $10$^{4}$. For flow in a circular
pipe this value of $\tau_w/\tau_w^{lam}$ is reached at $Re\simeq
10^{5}$, which is usually considered as a region of rather
developed turbulence \cite{landau,Tritt}.

Mechanical degradation of the polymers was quite small at shear
rates below 1.5 s$^{-1}$ and 1 s$^{-1}$, for $d=10$ mm and $20$
mm, respectively. The dependencies of $\tau_w/\tau_w^{lam}$ on
$\dot{\gamma}$ were therefore reproducible in consecutive runs
within about 1\%. At higher shear rates degradation effects became
appreciable. Thus, to ensure consistency of the results, time of
measurements at high shear rates had to be made possibly short.
So, for $\dot{\gamma}$ above 1.5 s$^{-1}$, curve 1, the shear rate
was raised by about 7\% per minute, and the measurement time was
not sufficient to average out fluctuations of $\tau_w$ produced by
the flow. Irregular undulations in curves 1 and 2 in Fig.3 at high
$\dot{\gamma}$ are due to those fluctuations and the short time of
averaging.

In spite of the short time of the experiments, the solution relaxation time, $%
\lambda$, typically decreased by up to 10\% after runs of the kind
shown by the curves 1 and 2. Nevertheless, a major part of the
degradation occurred at the highest shear rates, above 5 s$^{-1}$
and above 3.5 s$^{-1}$ for the curves 1 and 2, respectively.
Therefore, the data at lower shear rates can be regarded as quite
consistent.

Resistance is an integral characteristic of a flow, and its
fluctuations give some information about flow events on the scale
of the whole system. In order to get information about
characteristic frequencies of the flow on this integral scale, we
measured spectra of fluctuations of angular velocity of the upper
plate at constant applied torque, $T$. (AR-1000 is a stress
controlled rheometer, so it was more feasible to apply a constant
torque and to monitor rotation velocity, than to do it the other
way around.) Spectra of fluctuations of $\Omega$ at a few $T$
corresponding to different
average shear rates (all above the transition point, $\dot{\gamma}\simeq1$ s$%
^{-1}$) are presented in Fig.4. Root mean squares, RMS, of
fluctuations of the angular velocity, $\Omega_{rms}$, were about
4\% of its average values, $\bar{\Omega}$, for all $T$. Because of
the low rotation rates and low oscillation frequencies, inertial
effects associated with acceleration of
the shaft of the rheometer and of the liquid were always quite small. ($I%
\dot{\Omega}_{rms}$ was less than 1\% of
$(T/\bar{\Omega})\Omega_{rms}$, where $I$ is the moment of inertia
of the system, and $\dot{\Omega}_{rms}$ is RMS of the time
derivative of $\Omega$.) So, the spectra in Fig. 4 should reflect
time scales of fluctuations of viscous and elastic stresses in the
flow.

As the average shear rate is raised, the power of fluctuations,
$P$, increases and characteristic frequencies become higher,
Fig.4. The general form of the spectra remains very much the same,
however. There are no distinct peaks, except for instrumental
peaks at $f$, which are multiples of the average frequencies of
the upper plate rotation, $\bar{\Omega}/(2\pi)$. Further, each
spectrum has a broad region, spanning about a decade in
frequencies, where dependence of $P$ on $f$ is close to a power
law, $P\sim f^{-\alpha}$, with $\alpha\simeq 4.3$. (Some increase
in the power of fluctuations at higher $f$ is due to instrumental
noise, and it was also measured in laminar flow of the pure
solvent, curve 6 in Fig.4.) The power law dependence implies that
flow events on the integral scale of the system occur in this
whole region of frequencies. As it was argued above, excitation of
fluid motion in a broad continuous range of frequencies is one of
the characteristic features of turbulence.

\subsubsection{ Temporal and spatial spectra.}

Time spectra of fluctuations of flow velocity in a point at
different constant shear rates (all above the elastic instability
threshold) are shown in Fig.5. One component of the velocity in
the horizontal plane was measured with LDV in the center of the
set-up, where its average value was zero. The shear rates,
$\dot{\gamma}$, for the curves 1-5 were the same as the average
shear rates for the curves 1-5, respectively, in Fig.4. The
spectra in Fig.5 have the same general features as those in Fig.4.
Power of the fluctuations and their characteristic frequencies
increase with $\dot{\gamma}$, but the spectra remain very similar
in their appearance. In particular, the spectra do not have
distinct peaks, and do have broad regions of frequencies, where
the power of fluctuations decays according to a power law $P\sim
f^{-\alpha }$. (Flattening of the curves at high $f$ is due to
instrumental noise; see also discussion of Fig.16 below.) Again,
for each spectrum the power law decay region spans about an order
of magnitude in $f$, that implies excitation of the fluid motion
in the whole range of the corresponding temporal scales. One can
see that characteristic frequencies of fluctuations in a point and
at the integral scale(cf. Fig.4 and Fig.5) are rather close. The
exponent, $\alpha $, is about 3.5, which is rather close to the
exponent of 4.3 in Fig.4, and much larger than the Kolmogorov
exponent of 5/3, found for velocity spectra of high $Re$ inertial
turbulence \cite {landau,Tritt}.

A few representative snapshots of the flow viewed from below are
shown in Fig.6. In order to visualize the flow, the fluid was
seeded with light reflecting flakes (1\% of Kalliroscope liquid).
The upper plate was black. So, the bright regions are those, where
the flakes are oriented parallel to the upper plate. The patterns
of the polymer solution flow above the transition, at $Wi=6.5$,
Fig.6a-b, and at $Wi=13$, Fig.6c-e, look quite irregular and
exhibit structures of different sizes. One can see, however, that
the structures tend to have a spiral-like form, which is probably
imposed by the average azimuthal flow and circular symmetry of the
set-up. Further, there is a dark spot in the middle, which appears
in most of the snapshots. It corresponds to the center of a big
persistent toroidal vortex, which has dimensions of the whole
set-up (see also below). Direction of the vortical motion was
inwards near the upper plate, downwards near the center and
outwards near the lower plate. Average flow velocity along the
radial direction was measured by LDV in a few points, and the
results agreed quite well with presence of the big persistent
toroidal vortex. Flow of the pure solvent at the same shear rate
was completely laminar that can be learned from the snapshot in
Fig.6f. It looks quite uniform up to some variation along the
radial direction due to varying shear rate and inhomogeneity along
the circumference due to somewhat uneven (and directed)
illumination.

The visual impression of spatial disorder in the flow at high $Wi$
is confirmed by a more careful analysis. Figure 7 shows average
Fourier spectra of the brightness profiles at $Wi=13$ along the
diameter and along the circumference at $r=2d$. The spectra were
averaged over long series of flow pattern snapshots taken in
consecutive moments of time. The wavelength is measured in units
of $d$, so that the wave number, $k$, of unity corresponds to a
length of $2\pi d$. Amplitude of fluctuations of the brightness
along the circumference does not have any peaks and it exhibits a
power law decay, $A\sim k^{-1}$, over almost a decade in the wave
number domain. This implies that there are no selected wave
numbers in the azimuthal direction and the fluid motion is excited
in a broad range of spatial scales. (The plateau at high $k$ may
be due to restricted spatial resolution of the method of
visualization that was used.)

The spectrum taken along the diameter shows a general tendency of
power law decay in even broader range of wave numbers. However, it
apparently differs from the azimuthal spectrum by a series of
broad peaks, which may be due to the persistent toroidal vortex
discussed above. The average flow shear rate and the local gap
ratio, $d/r$, vary along the radius, so that the radial direction
is not neutral. Thus, the flow is not structureless and
homogeneous along the radial direction. The visualization of the
flow with the light reflecting flakes does not provide direct
information about the fluid velocity field. Therefore, the
specific value of the exponent in the power law fit in Fig.7 does
not have any special meaning.

In order to obtain quantitative information about spatial
structure of the fluctuating velocity field, we explored velocity
spectra in various off-center points, with non-zero average
azimuthal velocity, $\bar{V}_{\phi}$. Spectra of fluctuations of
the radial component of velocity, $V_r$, at $\dot{\gamma}=4$
s$^{-1}$ ($Wi=13.5$) at four different radii are shown in Fig.8.
They all were measured at $z=d/2$, where $z$ is the distance from
the upper plate. Root mean squares of the fluctuations in all four
points were rather close, varying between 0.88 and 0.99 mm/s. The
spectra in Fig.8 bear the same general features of turbulence as
the spectra in Figs.4, 5 and 7 discussed above. They do not
exhibit distinct peaks, and have broad regions of $f$, where the
power of fluctuations, $P$ decreases according to $P\sim
f^{-\alpha}$.

One can learn from Fig.8 that as the point of measurement is moved
away from the center, characteristic frequencies of the
fluctuations become higher. The most
reasonable explanation of that is growth of the average azimuthal velocity, $%
\bar{V}_{\phi }$, which was 3.81 mm/s and 6.99 mm/s, at $r=2d$ and
$r=3d$, respectively. So, the fluctuations of the velocity in time
in these two points are mainly due to fluctuations in space, which
are advected by the large mean flow velocity, $\bar{V}_{\phi }$.
Applying the Taylor hypothesis, we can view the spectra in time as
spectra in space, with relation between the frequency and the wave
number given by $k=2\pi f/\bar{V}_{\phi }$. Then the power law
decay regions in curves 3 and 4 imply that the fluid motion is
excited in the whole corresponding ranges of $k$. The ranges of
the spatial scales, where the motion is excited, span about an
order of magnitude for the both curves. The exponents $\alpha$ in
the power laws $P\sim f^{-\alpha }$ (and, so, $P\sim k^{-\alpha
}$) are again quite large, about 3.6 for $r=2d$ and about 3.3 for
$r=3d$. These large values of $\alpha $ imply that the power of
fluctuations decays very quickly as the size of eddies decreases.
The main contribution to fluctuations of both velocity and
velocity gradients (the power of the latter should scale as
$k^{-(\alpha -2)}$) should therefore be due to the largest eddies.

\subsubsection{ Mixing in the flow.}

Mixing in the flow was observed using a droplet of black ink added
to working fluid before rotation of the upper plate was started,
Fig. 9. Using a micropipette the droplet was carefully placed near
the center (at $r=0$) at about a half of the fluid depth. The ink
was dissolved in a concentrated sugar syrup, to match density of
the droplet with the density of the working fluid. We used the
half-size set-up with $R=19$ mm, $R_2=21.9$ mm, $d=5$ mm. The
volume of the working fluid was about 9 ml, and the amount of the
ink was 50 $\mu$l. The upper plate was colored white and the fluid
was viewed from below. So, regions without the ink appeared
bright, whereas dark regions corresponded to positions in the $(r,
\phi)$ plane, where the total amount of the ink across the fluid
layer was large.

Consecutive stages of mixing in the polymer solution are shown in
Fig.9. Rotation of the upper plate was abruptly started at $t=0$
at a rate of 1 turn per
4.26 seconds, which corresponded to a shear rate $\dot{\gamma}=5.6$ s$^{-1}$%
. It took about one minute for the irregular flow to develop,
after the rotation of the upper plate was started. (Development of
the irregular flow was judged by growth of the flow resistance,
which was saturated after about one minute, cf.\cite{stretch}.)
So, no significant changes in the ink distribution occurred during
the first 15 seconds (Fig.9). After 30 sec the ink became spread
over the surface of the lower plate by the big toroidal vortex
discussed above. In the snapshots taken at later moments of time
(60, 90 and 120 sec in Fig.9) one can see many fine scale
structures. Those can be either due to excitation of the fluid
motion at small spatial scales or significant stretching of fluid
elements along their Lagrangian trajectories by randomly
fluctuating large scale eddies. The contrast of the patterns
gradually decreases with the time that indicates progressing
mixing. The pattern in the last snapshot, taken 8 minutes after
the flow has been started, looks completely homogeneous. From the
appearance of the mixing patterns in Fig.9, characteristic time of
mixing can be estimated as 120 seconds, corresponding to about 30
full turns of the upper plate. The time required for mixing by
molecular diffusion without macroscopic flow can be estimated as
$dR/D$. Substituting for $D$ the diffusion coefficient for the
saccharose molecules, $D=8.5\cdot10^{-7}$ cm$^2$/s, we obtain a
diffusion time of about $10^6$ seconds, that is four orders of
magnitude larger than the mixing time in the flow. (Diffusion time
for suspended solid particles of the black ink could be
significantly larger.)

Mixing in the flow of the pure solvent at the same shear rate is
shown for comparison in Fig.10. One can see that distribution of
the ink remained inhomogeneous even after 9 hours, although the
ink became somewhat spread with the time. The Reynolds number was
about 0.5, and there were some non-vanishing inertial effects in
the flow. In particular, centrifugal force gave rise to a slow
fluid motion out of the $r-\phi$ plane. This motion had a form of
a big toroidal vortex with outflow near the upper plate and inflow
near the lower plate. (So, this vortex spun in the direction
opposite to that of the toroidal vortex produced by the elastic
stresses.) The vortical motion stirred the fluid but it did not
really mix it. So, when the set-up was viewed from a side, one
could see that the ink did not get to the toroidal core of the
vortex.

Summarizing the experimental results, we conclude that the flow of
the elastic polymer solution at sufficiently high $Wi$ has indeed
all the main features of developed turbulence that were stated
above. The fluid motion is excited in broad ranges of frequencies
and wave numbers, both spanning about an order of magnitude. The
flow is accompanied by dramatic increase in the rate of transfer
of momentum and mass. By the strength of the turbulent resistance
and by the span of scales in space and time, where the fluid
motion is excited, the observed flow can be compared to turbulence
of a Newtonian fluid in a pipe at $Re$ of about 10$^{5}$. This
apparently turbulent fluid motion in the swirling flow between two
plates arises at very low $Re$, where inertial effects are
negligible, solely because of the non-linear mechanical properties
of the elastic polymer solution. We therefore call the phenomenon
\textit{elastic turbulence}. Distinctions between the elastic
turbulence and the usual inertial turbulence, which is observed in
Newtonian fluids at high $Rey$, are discussed in Section III.D
below.

\subsection{ Further properties of the flow.}

\subsubsection{Flow structure and velocity profiles.}

In order to learn more about velocity field generated by the
elastic turbulence, we measured average velocity and RMS of the
velocity fluctuations at different points. Profiles in
$z$-direction of the average azimuthal velocity, $\bar{V}$, and of
RMS of its fluctuations, $V_{rms}$, at different flow conditions
are shown in Fig.11. (It was the standard set-up.) The
measurements are done at $r=2d$, that is rather far from the edge
of the upper plate ($R-r=1.8d$). Profile of $\bar{V}$ in a low
$Re$ flow of the pure solvent is an almost straight line, curve 3.
The elasticity driven turbulent flow significantly changes the
distribution of $\bar{V}$. It produces a high shear rate layer
near the upper plate, curve 1, and a low shear rate region near
the middle of the gap (at $z/d=0.5$). Such a distribution of
$\bar{V}$ is reminiscent of average velocity profiles in usual
high $Re$ turbulence. The perturbation of the $\bar{V}$ profile
becomes stronger when $\dot{\gamma}$ is increased, curve 2. In
particular, the slope of the $\bar{V}$-curve at small $z/d$
becomes larger that obviously corresponds to growth of $\tau
_{w}/\tau _{w}^{lam}$ with $\dot{\gamma}$ (Fig.3).

Fluctuations of the azimuthal velocity, curve 4 in Fig.11, are
small near the upper plate, reach a maximum at $z/d\simeq 0.25$,
and start to decrease at larger $z$. Again, such distribution of
$V_{rms}$ along $z$ is reminiscent of velocity fluctuations in
turbulent flows of Newtonian fluids \cite{landau,Tritt}. RMS of
fluctuations reaches a value of about $0.5d/\lambda $, so that
rate of deformations produced by the fluctuating velocity field is
on the order of $1/\lambda$. That implies an essentially
non-linear relation between the rate of deformations and the
fluctuating elastic stress. (It is discussed in more detail in the
next section.) The maximal RMS of the velocity fluctuations at
$\dot{\gamma}=4$ s$^{-1}$ was about 1.55 mm/s, that was about
7.5\% of the upper plate velocity ($V_{max}$) at $r=2d$ and about
25\% of the local value of $\bar{V}$.

As one can learn from curves 1 and 2 in Fig.11, there is
significant
asymmetry in distribution of the average shear rate, $\partial \bar{V}%
/\partial z$, along $z$. The main decrease of $\bar{V}$ occurs near the upper plate. So, at
$\dot{\gamma}=4$ s$^{-1}$, $\bar{V}/V_{max}$ becomes 0.5 at $z/d=0.12$ and at $z/d=0.5$ the average
velocity is only $0.18V_{max}$. (Velocity profile near the lower plate could not be measured because
of
instrumental restrictions.) The uneven distribution of the shear rate along $%
z$-direction should create an essentially inhomogeneous profile of $N_{1}$
along $z$. Indeed, in a laminar shear flow we have $N_{1}=2\eta _{p}\lambda \dot{%
\gamma}^{2}=2\tau _{p}\lambda \dot{\gamma}$ (where $\tau _{p}$ is
the polymer contribution to the shear stress). Therefore, the high
shear rate near the upper plate should generate large $N_{1}$ and
large hoop stress there. A result of this large unbalanced hoop
stress would be motion of the fluid near the upper plate inwards
in the radial direction, which is quite consistent with our
observation of the permanent toroidal vortex discussed above (see
Fig.6). In fact the shape of the $\bar{V}(z)$ profiles in Fig.11,
which is strongly asymmetric with respect to the mid-plane
($z=d/2$), is also consistent with the picture of a
\textit{single} large scale toroidal vortex. Indeed, even a
possible increase in the slope of the $\bar{V}(z)$ profile near
the lower plate still would not match the high shear at small $z$,
and the vortical motion in the $rz$-plane should be mainly driven
by the high hoop stress near the upper plate. In contrast, a
symmetric $\bar{V}(z)$ profile would imply equally high hoop
stress near the upper and the lower plate (and low hoop stress in
the middle) that would produce a pattern of two toroidal vortices
on top of each other.

In fact the turbulent $\bar{V}(z)$-profiles of the kind shown by
curves 1 and 2 in Fig.11 were not always stable, especially at
larger gap ratios. So, in a system with $d/R=0.563$ at
$\dot{\gamma}=2.5$ s$^{-1}$ the high shear rate layer randomly
moved between a vicinity of the upper plate and a vicinity of the
lower plate with a typical residence time of a few tens of minutes
and a transition time of about a minute. These fluctuations in the
location of the high shear rate layer were accompanied by
inversions of the
spinning direction of the big toroidal vortex. When the main drop of $\bar{V%
}$ occurred near the lower plate, it was its vicinity where the
radial flow was directed inwards. Therefore, the asymmetry of the
$\bar{V}(z)$ profile with respect to the $z=d/2$ plane (Fig.11) is
indeed directly related with the large scale vortical flow.

Big toroidal vortices driven by the hoop stress are actually quite
well known to appear in swirling flows of elastic fluids
\cite{bird,Boger2}, and inhomogeneity of the shear rate profile in
the primary laminar flow has long been recognized as their common
origin. In our system this vortex first arises as a stationary
structure at low shear rates. As it can be learned from Fig.3
(curve 1 at $\dot{\gamma}<0.75$ s$^{-1}$ and curve 2 at $\dot{\gamma}<0.4$ s$%
^{-1}$) the toroidal vortex leads to some increase in flow
resistance even before the transition to the elastic turbulence.
As it is suggested by the first three snapshots in Fig.9, the big
toroidal vortex is also the first vortical motion to arise upon
sudden inception of a shear rate, which is above the elastic
instability threshold level. (Further redistribution of the
average azimuthal velocity due to development of the turbulent
flow should enhance this large scale vortical motion.) So, we can
conclude that transition to the elastic turbulence in the swirling
flow between two plates is mediated by this vortex. The vortex is
always present in the flow, and it probably suffers permanent
instability and is permanently fluctuating, that may be an
important mechanism of production of fluid motion at smaller
scales. The toroidal vortex is providing a smooth, large scale
velocity field, which is randomly fluctuating in time, and in
which the liquid and the stress tensor imbedded in it are
chaotically advected. This type of advection can create variations
of stress in a range of smaller scales, which may cause small
scale fluid motion\cite{Volodya,Volodya2}. (This would be
analogous to generation of small scale concentration variations in
chaotic mixing by large fluctuating vortices.)

The large scale vortical flow is a probable reason for low $Wi$ at
the elastic instability threshold, and for the very strong
increase of resistance above the transition to elastic turbulence
in the flow between two plates. Indeed, growth of the flow
resistance in the CT flow at comparable $Wi$ (Fig.18 below) is
significantly smaller. This situation is somewhat similar to
inertial turbulence in Newtonian fluids in the same systems. In
the swirling flow between two plates a big toroidal vortex, which
is driven by centrifugal force, causes transition to turbulence at
quite low $Re$ and provides a route of efficient energy transfer
to the turbulent flow. Many instructive pictures of the toroidal
vortex, its evolution and instabilities in inertial,
inertia-elastic, and elasticity dominated flow regimes in a
swirling flow set-up with larger gap ratios can be found in
Ref.\cite{Boger2}.

\subsubsection{ Velocity statistics in a point.}

Characteristic probability distribution functions (PDF) of
azimuthal and radial components of the flow velocity in the regime
of elastic turbulence are shown in Fig.12a,b. These distributions
were obtained from LDV measurements taken at $r=2d$, $z=0.25d$ at
a shear rate of 4 s$^{-1}$. PDF's of the azimuthal and radial
velocity can be fitted quite well by the functions $A_{1}\cdot
exp[-(V-\bar{V}_{\phi })^{2}/(2a_{1}^{2})+0.032(V-\bar{V}_{\phi
})^{3}/a_{1}^{3}]$ and $A_{2}\cdot exp[-(V-\bar{V}%
_{r})^{2}/(2a_{2}^{2})-0.017(V-\bar{V}_{r})^{3}/a_{2}^{3}]$, respectively. Here $\bar{V}_{\phi }=6.34$
mm/s and $\bar{V}_{r}=-1.84$ mm/s are the average velocities; $a_{1}=1.55$ mm/s and $a_{2}=0.97$ mm/s
are the RMS of fluctuations for the azimuthal and radial components, respectively. One can see that
the both distributions have only minor skewness and are very close to Gaussians. (That skewness is
probably due to the non-zero average velocity.)

Probability distribution functions for gradients of velocity
(rates of deformation) in the longitudinal and transversal
directions, obtained from the same LDV time series, are shown in
Fig.13a and Fig.13b, respectively. The rates of deformation are
multiplied by the relaxation time, $\lambda $, to make them
dimensionless. The velocity gradients were estimated using the
Taylor hypothesis, with smoothing over a distance of about 1.45
mm. The difference, $\Delta V$, between consecutive velocity
readings with even time intervals of $\Delta t=0.22$ s was divided
by $\Delta t$ and by the average velocity $\bar{V}=(\bar{V}_{\phi
}^{2}+\bar{V}_{r}^{2})^{1/2}=6.6$
mm/s. One can see that the distributions of ${\frac{\partial V_{\phi }}{%
r\partial \phi }}$ and ${\frac{\partial V_{r}}{r\partial \phi }}$
can not be fitted by Gaussians. In contrast to the velocity
distributions in Fig.12, the both PDF's in Fig.13 have well
pronounced exponential tails, which imply significant
intermittency of the velocity gradients. The situation of nearly
Gaussian statistics of velocities and essentially non-Gaussian,
strongly intermittent distributions of velocity gradients is
actually quite typical for high $Re$ inertial
turbulence\cite{Frish}. Hence, the elastic turbulence resembles
high $Re$ inertial turbulent flows in this respect.

\subsection{Comparison with inertial turbulence at high Reynolds number.}

\subsubsection{ Role of the elastic stresses.}

Along with the apparent similarity in phenomenology between the
elastic and inertial turbulence, there are also quite a few very
important distinctions. An obvious reason for that is difference
in the physical mechanisms, which underlie the two kinds of
turbulent motion. Turbulence in Newtonian fluids is observed at
high $Re$, when the equation of motion (Eq. 1) is strongly
nonlinear. The nonlinear term in Eq. 1 is connected with
kinematics of \textit{translational} \textit{motion} in a
continuous medium and is due to fluid inertia. The elastic
turbulence occurs at low $Re$, when nonlinearity of the equation
of motion (Eq. 2) is negligible. The origin of the elastic
turbulence is nonlinear dependence of the elastic stresses on
rate of deformation in flow. (This nonlinear dependence is very
well seen in the model constitutive relation, Eqs. 3,4.) The
latter nonlinearity is connected with kinematics of
\textit{deformation} in a continuous medium, and it is due to
finite relaxation time of
 polymer molecules. Therefore, one can suggest that in the case of the elastic turbulence it would
be more relevant to study the field of stresses and of rates of deformations rather than the velocity
field.

The products of the randomly varying rates of deformation
(${\frac{\partial V_{\phi }}{r\partial \phi }}$ and
${\frac{\partial V_{r}}{r\partial \phi }}$ in Fig.13) and $\lambda
$ can actually be viewed as different estimates for Weissenberg
number, $Wi^{\prime }$, of the fluctuating velocity field. The
distributions in Fig.13 show that $Wi^{\prime }$ is rather often
above unity, and RMS of its fluctuations estimated from Fig.13a is
about 0.83. Fluctuations of the rate of deformation are directly
connected with fluctuations of the elastic stresses by the
constitutive relation (see Eqs. 3 and 4). Since the fluctuating
Weissenberg number is rather high, this connection should be
essentially nonlinear. Therefore, one can suggest that the field
of $\mathbf\tau _{p}$ in the elastic turbulence is highly complex
and strongly intermittent. It would certainly be quite instructive
to explore spatial structure and temporal distribution of the
elastic stresses, but there is currently no technique for local
measurements of $\mathbf\tau _{p}$ in a turbulent flow. On the
other hand, large scale properties of the $\mathbf\tau _{p}$ field
can be inferred from measurements of the total flow resistance.

As it is well known, the high flow resistance in high $Re$
inertial turbulence is due to large Reynolds stresses. The
Reynolds stress tensor is defined by the average values of $\rho
<V_{i}V_{j}>$, where $V_{i}$ and $V_{j}$ are different components
of flow velocity. In the case of elastic turbulence the Reynolds
stresses are quite small, since $Re$ is low. Indeed,
$<V_{i}V_{j}>$ can be estimated as $2V_{rms}^{2}$. Taking
$V_{rms}=1.55$ mm/s, which is the
maximal value for RMS of the azimuthal velocity fluctuations at $r=2d$ and $%
\dot{\gamma}=4$ s$^{-1}$, we obtain a Reynolds stress of $6\cdot
10^{-3}$ Pa, while $\tau _{w}$ at the same conditions is about 16
Pa. Therefore, the whole rise of flow resistance in the elastic
turbulence is caused by growth of average elastic stresses
\cite{Nat,stretch}. By the same token, the irregular fluctuations
of the total flow resistance in the elastic turbulence, that were
discussed above (Fig.4), are due to fluctuations in the
$\mathbf\tau _{p}$ field at large spatial scales.

Average growth of the polymer stresses can be estimated from the
curves in Fig.3. Indeed, by momentum conservation the stress near
the wall should be equal to its average value in the bulk, $\tau
_{w}=$ $\tau _{p}+\tau _{s}$, $\tau _{w}^{lam}=$ $\tau
_{p}^{lam}+\tau _{s}^{lam}$ (we omitted signs of averaging for
brevity). The viscous stress due to Newtonian solvent, $\tau_{s}$,
is linearly proportional to the shear rate, and its average value
depends only on the boundary conditions and can not change as a
result of a
turbulent flow \cite{stretch}. Therefore, we can take $\tau _{s}=\tau _{s}^{lam}=$ $\frac{\eta _{s}%
}{\eta -\eta _{s}}\tau _{p}^{lam}$, where $\eta $ is viscosity of
the polymer solution measured in a laminar shear flow at a
corresponding shear rate. Then for the ratio of the average
polymer stresses
in turbulent and laminar flow we obtain a relation $\frac{\tau _{p}}{\tau _{p}^{lam}}=\frac{%
\eta }{\eta -\eta _{s}}(\frac{\tau _{w}}{\tau
_{w}^{lam}}-\frac{\eta _{s}}{\eta })$. After plugging in numeric
values from Fig.3 (curve 2) and from measurements of $\eta
(\dot{\gamma})$ (Fig.2) we get ratios of $\frac{\tau _{p}}{\tau
_{p}^{lam}}$ as high as 65 at high shear rates. For a set-up with
larger $d/R$ and a solution with a smaller polymer concentration
the ratio of the polymer stresses in turbulent and laminar flows
estimated this way reached a factor of 170\cite{stretch}.

In molecular theory of polymer dynamics the tensor $\mathbf\tau
_{p}$ is found to be proportional to the polymer concentration and
to the average tensor of inertia of the polymer molecules, $\tau
_{p,ij}\sim k<R_{i}R_{j}>$ (where $k$ is an elasticity
coefficient, which remains constant in a linear elasticity
regime). Since elastic energy density of the stretched polymer
molecules is given by a quadratic form of a similar kind, one can
suggest linear dependence of the elastic energy on $\tau _{p}$.
Therefore, the density of the elastic energy should increase in
the turbulent flow by about the same factor as $\tau _{p}$. The
density of kinetic energy is proportional to the Reynolds stress
and although it certainly increases with development of the
elastic turbulence, it always remains quite small compared to the
elastic energy density. So, we can suggest that the elastic
stress, $\mathbf\tau _{p}$, should be the object of primary
importance and interest in the elasticity driven turbulent flow,
and it may be appropriate to view elastic turbulence as turbulence
of the $\mathbf\tau _{p}$ field.

The growth of the elastic stresses is also an evidence of significant extension of the polymer
molecules in the flow. The relation $\tau
_{p,ij}\sim k<R_{i}R_{j}>$ suggests that polymer extension should grow as $%
\sqrt{\tau _{p}}$. Taken together with the data on $\frac{\tau
_{p}}{\tau _{p}^{lam}}$, it implies an eight-fold stretching of
the polymer molecules in the turbulent flow beyond their extension
in a laminar flow with the same average shear rate. (For
$\frac{\tau _{p}}{\tau _{p}^{lam}}=170$ the corresponding
extension factor is 13.) Dramatic extension of the flexible
polymer molecules in the turbulent flow environment, inferred here
from the bulk measurements of the flow resistance, has been
recently confirmed by direct visualization of individual polymer
molecules in flow\cite{corinne}.

Behavior of polymer molecules in chaotic and turbulent flows has
been a subject of a long controversy\cite{Gyr}, and it deserves a
special discussion. Dynamics of polymer liquids have been most
well studied in basic
rheometric flows, such as simple shear and extension. In the former case $%
\tau _{p}$ is usually found to increase slower than linearly with the rate of deformation
$\dot{\gamma}$ (shear thinning). In the latter case $\tau _{p}$
grows faster than linearly with the rate of deformation, $\dot{\varepsilon}$%
, and at $\dot{\varepsilon}>1/(2\lambda )$  $\tau _{p}$ can grow by a few
orders of magnitude as the total deformation (measured by the Henky strain, $%
\int \dot{\varepsilon}dt$) increases. Molecular basis of this
behavior, stretching and tumbling in the case of a shear flow, and
a coil-stretch transition in the case of an extensional flow, has
been clearly demonstrated in the recent experiments by Chu's
group\cite{Chu-shear,Chu}. A turbulent flow involves complex
three-dimensional velocity field, with a tensor of rate of
deformation randomly varying in time and space. Therefore, an
attempt to directly fit it into the context of rheometric flows
may lead to frustration and confusion.

Being stochastic phenomena, chaotic and turbulent flows are most
appropriately treated statistically. An object of particular
interest for polymer dynamics is statistics of deformations of the
fluid elements, in which the polymer molecules reside. Since size
of the polymer molecules is almost always much smaller than the
size of the smallest turbulent eddies, velocity field is smooth on
the scale of the molecules. At every moment of time it is defined
by some average velocity and a tensor of rate of deformation, $\vec{V}(\vec{r%
})=\vec{V}_{0}+\frac{\partial V_{i}}{\partial
r_{j}}(\vec{r}-\vec{r}_{0})$. Statistics of deformations of a
fluid element can be derived from statistics of the tensor
$\frac{\partial V_{i}}{\partial r_{j}}$ in the
Lagrangian reference frame moving with the fluid element. The tensor $\frac{%
\partial V_{i}}{\partial r_{j}}$ is three-dimensional and in all
non-degenerate cases it has a positive eigenvalue, corresponding
to extensional flow along some axis. Rate of the extension and
direction of the axis vary randomly at a time scale, connected
with characteristic time of variations in the flow. A detailed
statistical analysis was first made by Lumley\cite{lumley} in
early 70's and was revised and refined
recently\cite{lebedev,chertkov}. Under quite general assumptions
about statistics of $\frac{\partial V_{i}}{\partial r_{j}}$, it
was shown that a fluid element, which is initially a sphere,
becomes an ellipsoid with its longest axis growing in average as
$a=a_{0}\exp (\alpha t)$. The coefficient $\alpha $ is a Lyapunov
exponent of divergence of two closely spaced points in the flow,
and the value of $\alpha $ is usually on the order of
$\left\langle \left( \frac{\partial V_{i}}{\partial r_{j}}\right)
^{2}\right\rangle ^{1/2}$. A somewhat surprising conclusion of the
theory is that a generic random flow is in average an extensional
flow in every point, with the rate of extension
$\dot{\varepsilon}=\alpha $ and unlimited Henky strain. Therefore,
one should expect the polymer molecules to become strongly
stretched by the random (turbulent) flow, when the Lyapunov
exponent $\alpha $, as given by the
statistics of $\frac{\partial V_{i}}{\partial r_{j}}$, becomes larger than $%
1/(2\lambda )$. (More refined theory suggests that a significant
fraction of the molecules should become substantially stretched
even at lower $\alpha $ due to intermittency in the
statistics\cite{lebedev,chertkov}.) One can see that the strong
extension of the polymer
molecules in the regime of elastic turbulence, taken together with the estimate $%
Wi^{\prime }\simeq 0.8$ are in good agreement with these theoretical
predictions\cite{lebedev,chertkov}.

Based on the discussion above, one can suggest the following
scenario of development of the elastic turbulence. The polymer
molecules are stretched in the primary shear flow that leads to
large elastic stresses, $\tau_{p}$. The elastic stresses render
the primary shear flow unstable and cause an irregular secondary
flow. The flow acts back on the polymer molecules stretching them
further and raising the elastic stresses even more\cite{stretch}.
That makes the flow increasingly turbulent, until a kind of a
saturated dynamic state is finally reached. This state implies
some mutually consistent fields of average stresses and
velocities, and their fluctuations. Those fields are connected
with each other by the equation of motion and the constitutive
relation.

\subsubsection{ Scaling with set-up size and fluid viscosity.}

We explored variation of different characteristics of the
elasticity induced turbulent flow with viscosity of the polymer
solution and with size of the set-up. It was in sharp
contradiction with the general principles of Newtonian fluid
mechanics, but quite in line with the concept of viscoelastic
similarity suggested for purely elastic flow transitions
\cite{EPL}. So, velocity required for excitation of inertial
turbulence in a Newtonian fluid is proportional to viscosity of
the fluid. This is connected with the fact that characteristic
inertial time (viscous diffusion time), $t_{vd}=d^2/\nu$, is
inversely proportional to viscosity. In contrast to it, polymer
relaxation time, $\lambda $, usually grows proportionally to
viscosity of the solvent\cite{bird}. When the system gap ratio and
the viscosity ratio are both kept constant, purely elastic
transitions should occur at a constant Weissenberg number,
$Wi=V\lambda /L$. Therefore, increase in viscosity of the solvent
should lead to transition to elastic turbulence at lower velocity.
In order to verify that we prepared PAAm solutions with the same
$\eta_p/\eta$ and a relaxation times larger than 3.4 s using a few
other sugar syrups with higher viscosities as solvents. The
transition to elastic turbulence in these solutions occurred at
the same $Wi$, indeed, while the angular velocity of the upper
plate and Reynolds number at the transition threshold were lower,
scaling as $\Omega\sim 1/\eta_s$ and $Re\sim 1/\eta_s^2$,
respectively. (In fact the situation was very similar to the case
of purely
elastic instabilities in CT flow \cite{EPL}.) The dependencies of $%
\tau_w/\tau_w^{lam}$ on $Wi$ for those solutions were close to
curve 2 in Fig.3. (In order to obtain a similar dependence,
though, it was important to vary the shear rate slower, in
proportion with the increase
in $\lambda$.) Using a very viscous sugar syrup with $\eta_s\simeq 7.2$ Pa$%
\cdot$s, we prepared a polymer solution with a relaxation time of
about 30 s, and observed a transition to elastic turbulence at
$\Omega\approx 0.05 s^{-1}$ and $Re\approx 3\cdot10^{-3}$
\cite{stretch}.

Further, the viscous diffusion time, $t_{vd}$, is proportional to
a square of the size of the system. Therefore, when size of the
set-up is proportionally reduced, the linear and angular
velocities required for
excitation of inertial turbulence should increase as $V\sim1/d$ and $%
\Omega\sim 1/d^2$, respectively. In contrast to it, polymer
relaxation time, $\lambda$, does not depend on system size.
Therefore, $\Omega$ (and $\dot{\gamma}$) at transition to the
elastic turbulence should not depend on size of the set-up, as
long as its proportions are preserved.
A natural scale of stress in the purely elastic flow regime is given by $%
\eta_p/\lambda$, and it also does not depend on size of the
system. (It may depend on $\eta_p/\eta_s$, though.) Therefore
transition to the elastic turbulence is expected to occur at the
same characteristic stress, and dependence of
$\tau_w/\tau_w^{lam}$ on $Wi$ is supposed to be the same for all
system sizes.

Figure 14 shows dependence of $\tau_w/\tau_w^{lam}$ on $Wi$ in a
set-up with all the dimensions proportionally reduced by a factor
of four compared with the standard configuration. The plot looks
quite similar to curve 1 in Fig.3. One can see that the transition
to the elastic turbulence indeed occurred at the same
$\dot{\gamma}$ as in the standard set-up, although the Reynolds
number was 16 times lower. In order to compensate for lower
sensitivity of the rheometer to shear stress due to smaller total
torque, the rate of variation of $\dot{\gamma}$ was reduced. That
may be the reason for a smaller region of hysteresis in Fig.14
compared with Fig.3. The plot in Fig.14 also implies that the
transition to elastic turbulence occurred at the same $\tau_w$ and
$Wi$, and dependence of $\tau_w$ on $\dot{\gamma}$ was also almost
the same as in the standard configuration.

The parameters of the system relevant to the elastic turbulence
are $d$, $R$, $\eta_p$, $\eta_s$, $\lambda$ and $\Omega$. (Density
of the fluid should be of no importance, since inertial effects
are negligible.) There seems to be no apparent way to construct
out of these parameters a scale of length other than the size of
the system (with a possible pre-factor depending on
$\eta_p/\eta_s$ and $\Omega \lambda$). It implies that, when the
dimensions of the system, $d$ and $R$, are proportionally changed,
while $\eta_p$, $\eta_s$, $\lambda$ and $\Omega$ are preserved,
the turbulent velocity field should remain the same up to linear
rescaling with size of the system.

This suggestion is qualitatively confirmed by the mixing patterns
presented in Fig.15. The snapshots in the left column are the same
as in Fig.9. They show consecutive stages of mixing of a droplet
of ink in a polymer solution upon sudden inception of shear flow.
The dimensions of the system were two times smaller than those of
the standard configuration. The snapshots in the right column show
mixing in the same polymer solution in a system with dimensions
reduced four-fold compared with the standard configuration. In
both cases the area of the snapshots corresponds to the area of
the upper plates. So, the snapshots in the right column are
magnified by a factor of two compared with the snapshots in the
left column. (The volume of the ink droplet added to the solution
in the right column was 4 times smaller than for the left column
in order to match the amount of ink per unit area.) The angular
velocities of the upper plates were the same, $\Omega =1.5$
s$^{-1}$, and so were the shear rates, $\dot{\gamma}=5.6$
s$^{-1}$, and the Weissenberg numbers. However, the Reynolds
number for the right column was four times smaller, $Re=0.125$
compared with $Re=0.5$ for the left column.

The mixing patterns in Fig.15 are irregular in the both columns,
but the two sequences look rather similar and make an impression
of two realizations of the same stochastic process. So the
transition to the elastic turbulence seems to go similarly in the
two systems, and the mixing seems to progress with about the same
characteristic time. This implies that characteristic velocities
of the irregular flow scale linearly with size of the system and
with velocity of the primary flow. Patterns, which appear in the
snapshots taken at equal elapsed times, seem to have similar sizes
and similar contrast levels. This suggests that in the elastic
turbulence characteristic spatial scales of velocity field are
indeed proportional to size of the system.

Quantitative data on dependence of the turbulent velocity field on
the system size are obtained from velocity spectra measured at a
point with a large mean flow velocity. As it was argued above,
those measurements provide information about both power of the
velocity fluctuations and spatial structure of the velocity field.
Spectra of fluctuations of the azimuthal and radial velocity
components measured at $r=2d$, $z=d/4$ in the standard and in the
half-size set-ups are shown in Fig.16. The shear rate was 4
s$^{-1}$ (corresponding to $Wi=13.5$) in the both systems, the
same as for the spectra in Fig.8 and for PDF's in Figs.12,13.

From the above similarity arguments fluctuating velocity was
expected to scale as $d/\lambda $. RMS of the fluctuations of
$V_{\phi}$ and $V_r$ were 1.55 mm/s and 0.97 mm/s, respectively,
for the standard configuration, and 0.78 mm/s and 0.48 mm/s,
respectively, for the half-size system. So the corresponding
fluctuation amplitudes in the half-size system were indeed twice
lower. Therefore, in order to match spectra in the two systems,
the velocity measured in the half-size system was multiplied by a
factor of 2. One can see, that upon this velocity rescaling, the
spectra of both $V_{\phi }$ and $V_{r}$ become very close. The
apparent difference at high frequency is due to lower
signal-to-noise ratio in the half-size system. Indeed, one of the
main sources of noise in the LDV measurements is finite size of
the crossing region of the laser beams, over which the measured
velocity is averaged. Characteristic velocity gradients (scaling
with $\Omega$ and $1/\lambda$) are expected to be equal in the two
systems, and dimensions of the laser beam crossing region are the
same as well. Therefore, the level of the gradient noise should
not change, when the size of the system is reduced, whereas the
level of the signal (characteristic fluctuating velocity)
decreases.

The average azimuthal velocities were 6.34 mm/s and 3.37 mm/s for
the standard and the half-size systems, respectively, with their
ratio being rather close to 2 again. Therefore, applying the
Taylor hypothesis to the power spectra in Fig.16, we conclude that
spatial structure of the velocity fluctuations in the two systems
is statistically identical up to rescaling by a factor of 2 in the
$k$-space.  Thus, the experimental results in Fig.16 provide
further, more quantitative support to the suggestion of simple
linear rescaling of the turbulent velocity field with size of the
system at constant $\Omega$ (and $\eta_p$, $\eta_s$ and
$\lambda$).

\subsubsection{ Flow at small scales and near walls.}

Structure of flow at small scales and near boundaries gives
another example of substantial differences between the elastic and
inertial turbulence. It is instructive to summarize some relevant
basic facts about the latter first. In high $Re$ inertial
turbulence the range of excited spatial scales is restricted from
below by the Kolmogorov
dissipation length, $l_{0}$. It can be estimated from the condition $%
Re_{l_{0}}\simeq 1$ \cite{landau,Frish}, with the local Reynolds
number, $Re_{l}$, defined as $V_{l}l/\nu $, where $V_{l}$ is a
characteristic turbulent velocity difference at the scale $l$.
Since $V_{l}$ decreases as $l$ becomes smaller (as $V_{l}\sim
l^{1/3}$ in Kolmogorov's picture of turbulence
\cite{landau,Frish}), $Re_{l}$ decreases faster than linearly with
decreasing $l$. So, the dissipation length $l_{0}$ necessarily
exists in any inertial turbulent flow.

An issue related to the problem of the smallest excited length
scale is stability of boundary layers near solid walls. The latter
can be most
simply approached by introduction of a local Reynolds number, $Re(z)=\bar{V}%
(z)z/\nu $, which is a growing function of distance, $z$, from the
wall. (The usual convention is to use $y$ for distance from a wall
(transverse direction), and $z$ as a coordinate in the spanwise
direction, but we would like to stay consistent with the notation
used in Fig.11 and cylindrical geometry.) Near a wall the mean
flow velocity, $\bar{V}(z)$, increases linearly with $z$, giving
$Re(z)\sim z^{2}$ at small $z$. Therefore, $Re(z)$ remains low in
some vicinity of a wall and the flow there is a laminar shear.
This laminar shear layer near a wall (usually called viscous
sublayer) ends at a distance $z$ (decreasing with growth of global
$Re$), where $Re(z)$ becomes sufficiently large, and the shear
flow becomes unstable and turbulent.

Analogously to the inertial turbulence, we can introduce a local
Weissenberg number, $Wi_{l}=\lambda V_{l}/l$. It does not decrease
as $l$ becomes smaller, though. Indeed, the ratio $V_{l}/l$ is a
characteristic gradient of the turbulent velocity smoothed over
the length $l$, and it can only grow as $l$ decreases. Since
$Wi_{l}$ does not get small at small $l$, there appears to be no
straightforward way to construct an analog of the dissipation
length, $l_{0}$, for the elastic turbulence. So, there may be no
macroscopic length setting a lower limit to the range of scales of
the turbulent motion. (Something may change, though, at a scale
comparable with size of the polymer molecules or with their
characteristic diffusion length.)

This suggestion is consistent with the plots shown in Fig.16.
Indeed, the spectra of both $V_{\phi }$ and $V_{r}$ in Fig.16 have
broad frequency ranges where the power of fluctuations decays as
$P\sim f^{-\alpha }$. The value of $\alpha $ is close to the
values found for the other velocity spectra (Fig.5, 8). The
highest frequency in a power law decay region corresponds to the
smallest spatial (or temporal) scale, at which the turbulent flow
is excited. The power law decay regions in the spectra in Fig.16
are cut off at high frequencies by plateaus rather than by abrupt
drops, however. As we argued before, those plateaus appear solely
due to the instrumental noise. The power law decay range and the
apparent range of excited spatial scales deduced from it become
broader, when the signal-to-noise ratio becomes higher, Fig.16.
This lack of the cut-off frequency from above does not lead to any
kind of high frequency divergence (ultraviolet catastrophe),
though. Indeed, since $\alpha>3$, the total power of the velocity
fluctuations is converging at high $f$ and $k$, and the same is
true for fluctuations of velocity gradients with their power
scaling as $P\sim k^{-(\alpha -2)\text{ }}$in the spatial domain.
(We assume again that the Taylor hypothesis is applicable.)

Similarly to $Re(z)$, one can introduce a local Weissenberg number
near a wall, $Wi(z)=\lambda \bar{V}(z)/z$ ($\bar{V}(z)$ here is
the difference between the local mean flow velocity and the
velocity at the wall at $z=0$), and easily see that it does not
become small at $z\rightarrow 0$. In fact, we can learn from the
average velocity profiles in the elastic turbulence, Fig.11, that
the region near the upper plate corresponds to the highest average
shear rate and to the highest $Wi(z)$. On the other hand, the
velocity fluctuations clearly decline near the upper plate,
Fig.11, and existence of such a stable high shear rate boundary
layer in the elastic turbulence may appear rather puzzling.

A possible mechanism providing stability of the flow near the
upper plate is significant shear thinning of the relaxation time,
$\lambda $. As it was argued above, in the elastic regime flow
stability depends on the parameter $K={\frac{\eta _{p}}{\eta
}}{\frac{d}{R}}Wi^{2}$. Adopting it for the boundary layer, we can
suggest that stability should be defined by a local parameter
$K(z)={\frac{\eta _{p}}{\eta }}{\frac{z}{R}}Wi^{2}(z)$, where
$Wi(z)=\lambda ({\dot{\gamma})}\bar{V}(z)/z$,
$\lambda({\dot{\gamma}})$ is the shear rate dependent relaxation
time, and $R$ is the radial position of the region of interest. We
can consider a case of a simple sigmoidal profile, where the whole
growth of the mean flow velocity from zero to $\bar{V}$ occurs
across thin layers of widths $z_b$ near the boundaries. If
$\bar{V}$, ${\frac{\eta _{p}}{\eta }}$, and $R$ are all kept
constant, we obtain $K(z_b)\sim \lambda ^{2}({\dot{\gamma}})/z_b$,
where ${\dot{\gamma}=}\bar{V}/z_b$. Therefore, if the relaxation
time decreases with the shear rate faster than $\lambda \sim
\dot{\gamma}^{-1/2}$, $K(z_b)$ is a decreasing function, and the
boundary layer remains stable as it becomes thinner and the slope,
$\bar{V}/z_b$, increases.

Such fast decay of $\lambda $ with $\dot{\gamma}$ would also
suggest that a more global (although radius dependent) parameter
for the
flow between two plates $K={\frac{\eta _{p}}{\eta }}{\frac{d}{R}[}\lambda ({%
\frac{\Omega R}{d})\frac{\Omega R}{d}]}^{2}$ would be decreasing
with $R$, and the elastic instability would be limited to a region
of sufficiently small $R$. This kind of behavior was indeed
reported for a different polymer solution and a system with small
$d/R$\cite{McKinley}. In a more general case of shear thinning,
with the relaxation time scaling as $\lambda
=\lambda _{0}\dot{\gamma}^{-\alpha }$ (and $\alpha <1/2$), the condition $%
K<K_{c}$ (where $K_{c}$ is a critical value for instability) is
satisfied for $\bar{V}(z)<\left[ \frac{K_{c}\eta R}{\eta
_{p}\lambda _{0}^{2}}\right] ^{1/(2-2\alpha )}\cdot
z^{\frac{1-2\alpha }{2-2\alpha }}$, which is reminiscent of the
$\bar{V}(z)$ profiles in Fig.11.

Dependence of $\lambda $ on $\dot{\gamma}$\ in a shear flow can be inferred
from measurements of normal stress difference, $N_{1}$. At $\lambda \dot{%
\gamma}\ll 1$ it is usually found that $N_{1}=2\lambda \eta _{p}\dot{\gamma}%
^{2}$ with constant $\lambda $ and $\eta _{p}$.\ Therefore, $\lambda (\dot{%
\gamma})=N_{1}/(2\eta _{p}\dot{\gamma}^{2})$ can be used as an estimate of
shear thinned relaxation time at higher $\dot{\gamma}$. Measuring this way $%
\lambda (\dot{\gamma})$ for solutions of a few lower molecular weight PAAm
samples we usually found $\lambda \sim \dot{\gamma}^{-\alpha }$, with $%
\alpha $ around 0.3 \cite{Ours2}. (Resolution of our rheometer was
not sufficient to measure normal stresses in the solution used in
the experiments on the elastic turbulence at $Wi$ in the range of
interest\cite {Ours2}.) Therefore the shear thinning may indeed
play an important role in stabilization of the boundary layer.

Finally, it is worth noting that the whole concept of boundary
layers is less straightforward in the case of elastic turbulence,
and the condition of their stability may be less restrictive than
for inertial turbulence. Indeed, in a high $Re$ turbulent flow
characteristic fluctuating velocity, $V'(z)$, generated by a flow
instability originating at a distance $z$ from a wall scales as
$V'(z)\sim\nu /z$, and the boundary layer must become stable at
some $z$, since $V'(z)\rightarrow \infty $ as $z\rightarrow 0$. In
contrast, characteristic fluctuating velocity generated by an
elastic instability should be proportional to $z/\lambda $ and
approaching zero at $z\rightarrow 0$.

\section{Couette-Taylor flow.}

\subsection{ Experimental set-up and results.}

The experiments on the Couette-Taylor flow were carried out in a
small set-up mounted vertically on the AR-1000 rheometer. The
stationary outer cylinder was made of glass, had a radius $R_{2}=$
13.5 mm, was open from above and closed from below by a stainless
steel plug with a flat top, Fig.17. The rotating inner cylinder
had a radius $R_{1}=$ 9 mm and was conical at the bottom with an
angle of $\tan ^{-1}(0.5)$, so that the gap ratio in the
cone-and-plate flow at the bottom was the same as in the annular gap, $%
d/R_{1}\equiv (R_{2}-R_{1})/R_{1}=0.5$. The cylindrical surface of
the inner cylinder was 72 mm long. The both cylinders were coaxial
with the shaft of the rheometer, and the inner cylinder was
attached to it. The working liquid was filled above the upper rim
of the inner cylinder. In order to reduce evaporation of the fluid
from its open surface a special cover (vapor trap) was used. The
outer cylinder was enclosed in a thermal jacket with a square
cross-section that was made out of plexiglass. The temperature of
the system was set by a fast stream of water pumped through the
jacket from a refrigerated circulating bath. The polymer solution
was the same as in the experiments on the flow between two plates
(80 ppm of PAAm in 65\% sucrose, 1\% NaCl in water), and the
experiments were carried our at the same temperature of 12
$^{\circ }$C. The characteristic shear rate in the flow was
calculated as $\dot{\gamma}=\Omega R_{1}/d$, and the Weissenberg
number was taken to be $Wi=\lambda \Omega R_{1}/d$ with $\lambda
=3.4$ s.

First, we measured flow resistance in the polymer solution as a
function of  $\dot{\gamma}$ and $Wi$. The protocol was very
similar to that described above for the flow between two plates.
At the beginning, pure solvent (with a viscosity $\eta_s$) was
filled into the system and torque, $T_s$, applied to the inner
cylinder was measured as a function of angular velocity, $\Omega
$. The dependence was linear over the whole range of $\Omega $
explored. Suggested dependence of torque for an (imaginary)
laminar flow of the polymer solution was calculated from it as
$T_{lam} = T_s \eta(\dot{\gamma})/\eta_s$. Then the system was
filled with the polymer solution, $\Omega $ was gradually
increased starting from a low point corresponding to $Wi=0.68$,
and the torque, $T$, required to drive the actual polymer solution
flow was measured. The ratio of the two torques, $T/T_{lam}$, is
equal to the ratio of the average stresses, $\tau /\tau _{lam}$,
in the actual and the imaginary laminar flow of the polymer
solution. It is a quantitative expression of growth of the flow
resistance compared with the laminar flow. The dependence of $\tau
/\tau _{lam}$ on $Wi$ is shown in Fig.18 (thin black line) for
$Wi$ varying over almost two orders of magnitude. One can see that
the resistance ratio $\tau /\tau _{lam}$ is equal to unity at low
values of $Wi$ corresponding to laminar flow in the system. Then
at $Wi$ of about 4 an abrupt transition occurs, and the flow
resistance starts growing quickly and non-linearly. Reynolds
number at the transition point is quite small, $Re=\Omega
R_{1}d\rho /\eta =0.07$. Therefore, the transition is of a purely
elastic nature. The non-linear growth of the flow resistance is
quite significant; at the highest tested value of $Wi$ the ratio
$\tau /\tau _{lam}$ is more than 4. Undulations of the curve in
Fig.18 reflect fluctuating nature of the flow resistance above the
transition and are due to limited integration time during the
measurements.

In order to explore behavior of the flow resistance at decreasing
shear rate, we carried out another experiment, where $Wi$ was
first set to 7, well above the transition point, and gradually
reduced afterwards. Dependence of $\tau /\tau _{lam}$ on $Wi$\ in
this run is shown by a thick gray curve in Fig.18. The two curves
in Fig.18 overlap at $Wi$ above 4.7 and below 1.8, whereas in the
region in between a well pronounced hysteresis occurs. This broad
region of hysteresis agrees quite well with previously reported
results for Couette-Taylor flow in a purely elastic regime, which
were obtained in a system with a smaller gap ratio and a different
PAAm solution\cite{Ours2,EPL}. The appearance of the plot in
Fig.18 is quite similar to the dependencies of $\tau /\tau _{lam}$
on \ $Wi$ for the flow between two plates shown in Figs.3 and 14.

The non-linear velocity transition was accompanied by emergence of
a disordered and randomly fluctuating pattern of vortices on top
of the basic shear flow. When visualized with the light reflecting
Kalliroscope flakes, the fluctuating pattern looked similar to the
Disordered oscillations reported before at similar flow conditions
in a larger CT system with $d/R_{1}=0.2$\cite{Ours2}. In order to
measure velocity associated with the fluctuating flow we used the
LDV set-up again. Two horizontal laser beams were crossing inside
the polymer solution seeded with polystyrene beads in the middle
of the annular gap at $ r=(R_{1}+R_{2})/2$ and at about half
height from the bottom. The beams were adjusted in such a way that
it was the radial component of the flow velocity, $V_{r}$,
perpendicular to the basic shear flow that was measured. Power
spectra of fluctuations of $V_{r}$\ at a few different $Wi$ above
the non-linear transition are shown in Fig.19. As $Wi$ is raised
the total power of the velocity fluctuations increases and
fluctuations at higher frequencies are excited. Nevertheless, all
four curves share the same general appearance: there are two
plateau regions at low and high $f$ (the latter should be due to
the instrumental noise) and two contiguous regions of a power law
decay in between, with two different exponents and a distinct
inflection point separating them. Although characteristic
frequencies increase with $Wi$, those exponents are rather close
for all four curves, around -1.1 at lower frequencies and around
-2.2 in the high frequency power law decay regions. Apart from the
inflection points, none of the curves exhibits any distinct peaks.
Amplitude of the velocity fluctuations in all four cases was quite
small compared with average flow velocity. The latter was directed
along the circumference (azimuthally) and equal to about $\Omega
R_{1}/2$. (See also Fig.20 below and discussion thereof).
Therefore, the spectra in Fig.19 should actually reflect spatial
variation of $V_{r}$\ in the azimuthal direction.

In order to obtain additional information about the fluctuating
flow velocity, we used the same time series to calculate velocity
statistics. The probability distribution functions of $V_{r}$\ at
different $Wi$ are shown in Fig.20. One immediately observes that,
although the average velocity is always zero, the shapes of the
distributions are essentially non-Gaussian. Moreover, unlike the
velocity distributions for the flow between two plates in Fig.12,
PDF's in Fig.20 cannot be satisfactory
fitted by skewed Gaussians (functions of the form $A\cdot exp[-(V-\bar{V}%
_{r})^{2}/(2a^{2})+b(V-\bar{V}_{r})^{3}/a^{3}]$ ). One of the most
notable features of the PDF's in Fig.20 is relatively high
probability
for large negative values of $V_{r}$ (''long tails'' of the distributions at negative $%
V_{r}$), which implies asymmetry for radial flows inwards and
outwards. Although most striking in the curve for $Wi=5.4$, this
feature is still quite well expressed at $Wi=35$. In fact, the
curve at $Wi=5.4$ is also quite similar to a PDF\ of $V_{r}$\
reported before in the CT flow with $d/R_{1}=0.2$,
Ref.\cite{Ours2} (Fig.12 therein), at comparable $Wi$ in the
regime of Disordered oscillations. In that latter case the
asymmetry in the radial velocity distribution was due to abundance
of coherent structures having a shape of couples of vortices
(''diwhirls'') with a narrow region of intensive radial flow
inwards in the middle and slow radial flow directed outwards on
the sides of a vortex couple. Such an asymmetric shape of the
diwhirls is closely related to mechanism of generation of the
elastic instability in CT flow, and it allows efficient pumping of
energy from the primary shear flow into the secondary fluctuating
velocity field \cite {Ours2}. Therefore, the asymmetric shapes of
PDF's in Fig.20 suggest that the coherent structures of that
diwhirl type are present in the CT flow even at high $Wi$ and may
have significant influence on its organization and statistical
properties.

The distribution functions in Fig.20 also allow finding
characteristic rates of extension in the flow in the radial
direction, $\partial V_{r}/\partial r $, and Weissenberg numbers,
$Wi^{\prime }=\lambda \partial
V_{r}/\partial r$, associated with them. Indeed, we have got boundary conditions of $V_{r}=0$ and $%
\partial V_{r}/\partial r=0$ at $r=R_{1}$ and $r=R_{2}$. Therefore, $\partial V_{r}/\partial r$
in the bulk can be estimated as $4V_{r}^{rms}/d$, where
$V_{r}^{rms}$ is RMS of the radial velocity fluctuations. The
resulting values of $\partial V_{r}/\partial r$ calculated from
PDF's in Fig.20 are then 0.059, 0.071, 0.097 and 0.15 $s^{-1}$ at
$Wi$ of 5.4, 13.6, 21.8 and 35, respectively. Those values of
$\partial V_{r}/\partial r$\ correspond to $Wi^{\prime }$ of 0.2,
0.24, 0.33 and 0.52, respectively. These estimated numbers are
rather high and are generally consistent with the picture of
strong elongation of the polymer molecules by elements of
extensional flow in the fluctuating velocity
field\cite{corinne,lumley,lebedev,chertkov}.

\subsection{ Discussion and comparison with the flow between two plates.}

Summarizing the results shown in Figs.18-20, we conclude that in
the purely elastic regime the CT flow behaves similarly to the
flow between two plates described in section III. The purely
elastic non-linear flow transition occurs at $Wi=O(1)$ and
(arbitrarily) small $Re$ in the both cases; the transition is
rather sharp, has pronounced hysteresis and it is accompanied by
significant growth of flow resistance. The fluctuating velocity
field above the transition is quite random for the both flows and
its spectra do not exhibit any distinct peaks, cf. Fig.8 and
Fig.19. Therefore, although characterized much less extensively,
the CT flow above the non-linear transition fits quite well the
definition of elastic turbulence introduced above.

Nevertheless, the elasticity induced turbulent flow in the CT
system differed from the turbulent flow between two plates
quantitatively, and the turbulent flow effects in the CT system
were less dramatic. Indeed, the non-linear transition in the CT
system occurred at $Wi\approx 4$ (Fig.18), which is
higher than critical $Wi$ in the flow between two plates, 3.2 and 2, for $%
d/R=0.26$ and $0.52$, respectively (Fig.3). The growth of flow
resistance in the CT flow above the transition was notably smaller
than in the flow between two plates - factor of 4.2 (Fig.18) vs.
factors of 11 and 19 (Fig.3) at the highest $Wi$ explored. (The
ratio $\tau /\tau _{lam}$ in the CT flow kept growing with $Wi$,
though, and it was the increasingly fast mechanical degradation of
the polymers that prevented us from raising $Wi$ higher, Fig.18.)
Characteristic amplitudes of fluctuating velocity in the CT flow
(Fig.19) were significantly smaller than in the flow between two
plates (Figs.11, 12), both in terms of $d/\lambda $ (0.06 vs. 0.5
at $Wi\simeq 13$) and of the maximal azimuthal flow velocity
(0.005 vs. 0.075 at $Wi\simeq 13$). This comparison is certainly
not quite fair, since $V_{r}$ in the CT flow is a transversal
velocity component. Hence, it should be best compared with $V_{z}$
in the flow between two plates, that was not measured. The general
tendency appears to be rather clear, though.

Characteristic estimated rate of extension due to the fluctuating
velocity field in the CT flow was also considerably smaller than
in the flow between two plates, $0.24/\lambda $ vs. $0.83/\lambda
$ at $Wi\simeq 13$. (We note here again, though, that these
estimates were obtained from different types of data sets, Fig.20
vs. Fig.13). Nevertheless, even the relatively low value of the
rate of extension in the CT flow is reasonably consistent with the
general condition for it to be on the order of $1/\lambda $\ to
cause significant elongation of the polymer molecules, growth of
the elastic stresses and notable flow resistance increase.

As it was already suggested above, more favorable conditions for
development of the elastic turbulence and stronger turbulent
effects in the flow between two plates may be due to the large
toroidal vortex. This vortex, which is driven by the distribution
of normal stresses in the basic shear flow, may provide a major
channel for pumping of energy into the turbulent velocity field.
CT flow certainly has a more limiting geometry, without any
coherent three-dimensional large scale flow of comparable
magnitude and with fluid motion confined to the annular gap. That
latter confinement limits possible curvature of flow trajectories
and may also make the general requirement of axial symmetry for
excited vortex structures (as dictated by the system geometry)
more severe and restrictive, somewhat impeding development of the
turbulence. In fact, axial symmetry of the vortex patterns is a
quite pronounced feature in inertial turbulence in CT flow even at
very high $Re$ \cite{Swinney}.

Among all the discrepancies between the elastic turbulence in the
CT system and in the flow between two plates, the most principal
one appears to be the difference in the decay exponents in the
power spectra of velocity fluctuations, cf. Figs.5, 8, 16 vs.
Fig.19. In the case of the flow between two plates the decay
exponents were around -3.5 in the both spatial and temporal
domains, while the power spectra in the CT flow had two distinct
power law decay regions with significantly lower exponents of -1.1
and -2.2. As it is shown in Section V below (Fig.25), velocity
spectra in an elastic turbulent flow in a curvilinear channel have
regions of a power law decay as well.  The decay exponents are
about -3.3, suggesting that it is the CT flow that behaves
''abnormally''.

A clue to the behavior of the CT flow may be the inflection points
in the spectra in Fig.19, which appear at frequencies
$f_{c}=(0.45\pm .05)\Omega /(2\pi )$ at all four $Wi$. These
frequencies are close to average rates of rotation (in revolutions
per second) of the fluid in the measurement point. Together with
the non-Gaussian and asymmetric shapes of PDF's in Fig.20, that
brings us to a suggestion that the power spectra in Fig.19 are
significantly influenced by the axisymmetric coherent structures
(possibly of the diwhirl type) discussed above. Those coherent
structures may dominate the low frequency part of the spectra,
$f<f_{c}$, and also feed into the high frequency part, $f>f_{c}$,
(where the ''original'' power of fluctuations should be rather low
to begin with, Figs.8, 16) reducing the apparent absolute values
of the decay exponents at $f>f_{c}$.

\section{ Flow in a curvilinear channel.}

\subsection{ Experimental set-up and procedure.}

The main motivation of the experiments on flow in a curvilinear
channel (Dean flow) was to carry out detailed quantitative study
of mixing in the elastic turbulence \cite{mix}. It is an open flow
that allows extended continuous experimental runs with
reproducible and well controlled initial conditions, and easy
gathering of extensive data at different stages of mixing. The
channel is schematically shown in Fig.21. It had a uniform depth
$d=3$ mm, was machined in a transparent bar of perspex, and it was
sealed from above by a transparent window. The channel consisted
of a sequence of smoothly connected half-rings with the inner and
outer radii $R_{1}=3$ mm and $R_{2}=6$ mm, respectively; it was
square in its cross-section, and had quite a high gap ratio,
$d/R_{1}=1$. The high gap ratio was intended to facilitate
development of an elastic instability at low $Wi$ and of intensive
irregular flow above the instability threshold. The channel had 30
repeating units, 18 mm in length each, Fig.21.

The liquids to be mixed were fed into the channel by two identical
syringe pumps through two separate tubing lines, always at equal
discharge rates. Chemical composition of the two liquids was
always identical as well, with the only difference of a small
concentration, $c_{0}=2$ ppm, of a fluorescent dye (fluorescein)
added to one of them. They were prepared from the same stock of a
carefully filtered liquid, which was divided into two equal parts.
A small amount of a concentrated solution of the dye was added to
one part, while the other part was diluted by an equal amount of
pure water. (Each of the liquids was carefully mixed afterwards.)
This method of preparation provided very good matching of
densities and refraction indices of the liquids. The channel was
illuminated from a side by an Argon-Ion laser beam converted by
two cylindrical lenses to a broad sheet of light with a thickness
of about 40 $\mu $m in the region of observation. It produced a
thin cut of the three-dimensional mixing pattern, parallel to the
top and the bottom of the channel at a half of the channel depth.

Fluorescent light emitted by the liquid in the direction
perpendicular to the beam and to the channel plane was projected
onto a CCD array by a camera lens and digitized by a 8-bit,
512$\times $512 frame grabber. Using homogeneous solutions with
different amounts of the dye, we found intensity of the
fluorescent light captured by the camera to be linearly
proportional to the dye concentration. Therefore, concentration of
the dye was evaluated from the fluorescent light intensity. (The
liquid without the dye appeared completely dark, Fig. 22, and
signal form it was below noise level of the camera.)

The flow was always observed near the middle of a half-ring on the
side, from which the laser beam was coming. So, the number, $N$,
of a unit (Fig.21) counted from the inlet was a natural linear
coordinate along the channel. Since the total rate of liquid
discharge, $Q$, was constant, $N$ was also proportional to the
average time of mixing. In order to observe further stages of
mixing (corresponding to $N>30$), we carried out a series of
experiments, where the liquids were pre-mixed before they entered
the channel. A shorter channel of the same shape was designed for
this purpose and put before the entrance to the original one. In
multiple calibration experiments we found good matching between
states of mixing at $N=2$ with the pre-mixer and at $N=27$ without
it\cite{mix}. Therefore, for the experiments with the pre-mixing a
number of 25 was added to the actual coordinate along the channel
to calculate the effective $N$.

Flow velocity was measured directly with the aid of the LDV
set-up. Because of small width of the channel, a special effort
was made to obtain high spatial resolution by reduction of the
region of space, where the two laser beams crossed, and reduction
of distance between the interference fringes. Focusing lenses with
a small focal length (about 25 mm) were used, and the angle
between the beams was raised to about 90$^{\circ }$ in the air
(and about 60$^{\circ }$ in the liquid). As a result, the region
of the beam crossing was brought down to $15\times 15\times 40$
$\mu $m, and the distance between the fringes was 0.44 $\mu $m.

We used the same polymer solution as in the experiments on the flow between two plates and on the CT
flow. (It was 80 ppm of PAAm, 65\% sugar, 1\% NaCl in water.) This time, however, the experiments were
made at room temperature, $22.5\pm 0.5$ $^{\circ }$C. So, the solvent viscosity was $\eta _{s}=0.153$
 $Pa\cdot s$, and viscosity of the solution was $\eta =0.198$ $Pa\cdot s$ at a
shear rate $\dot{\gamma}=4$ s$^{-1}$. The relaxation time,
$\lambda $, estimated from phase shift between the stress and the
shear rate in oscillatory tests with a shear rate amplitude of 3
$s^{-1}$, was 1.4 s. An estimate for the diffusion coefficient of
the dye was given by that for the saccharose molecules,
$D=8.5\cdot 10^{-7}$ cm$^{2}$/s. The characteristic shear rate and
the Weissenberg number in the flow were estimated as $\dot{\gamma}
=(2Q/d^{2})/(d/2)=4Q/d^{3}$ and $Wi=\lambda (4Q/d^{3})$,
respectively. The Reynolds number was calculated as $Re=2Q\rho
/(d\eta )$.

\subsection{Results.}

The Reynolds number in the flow was always quite small, reaching
only 0.6 for the highest $Q$ that we explored. Therefore, flow of
the pure solvent always remained laminar and no mixing occurred,
Fig.22a. The boundary separating the liquid with and without the
dye was smooth and parallel to direction of the flow and only
became somewhat smeared due to molecular diffusion as the liquid
advanced downstream. Flow of the polymer solution was laminar and
stationary up to a value of $Q$ corresponding to $Wi_{c}=3.2$ (and
$Re=0.06$), at which an elastic instability occurred. This
instability lead to irregular flow and fast mixing of the liquids.

A few typical mixing patterns at different $N$ in the polymer
solution above the instability threshold are shown in photographs
in Fig.22b-d. More insight about structure and evolution of the
mixing patterns can be obtained from space-time diagrams.
Representative diagrams taken at $Wi=6.7$ at four different $N$
are shown in Fig.23. Brightness profiles along a single line going
perpendicularly to the channel near the middle of a half-ring (a
horizontal line going through the middle of a snapshot in Fig.22)
were captured at even time intervals of 80 ms and plotted from top
to bottom. The diagrams in Fig.23 share the same chaotic
appearance and show features at comparable scales, but they loose
contrast as liquid advances downstream and gets progressively
mixed.

As it is illustrated by the space-time diagrams in Fig.23, mixing
in the polymer solution flow above the instability threshold was a
random process calling for statistical analysis\cite{mix}. A
simple parameter characterizing homogeneity of the mixture is a
root mean square of deviations of the dye concentration from its
average value, $\bar{c}=c_{0}/2$, divided by the average value
itself, $c_{rms}=<(c-\bar{c})^{2}>^{1/2}/\bar{c}$. Small value of
$c_{rms}$ means high homogeneity and good mixing of the liquids.
At the channel entrance, where the two injected liquids are
perfectly separated, $c_{rms}$ is unity, and it should become zero
for a perfectly mixed liquid.

Dependence of $c_{rms}$ on $Wi$ near the exit of the channel, at
$N=29$, is presented in Fig.24. Statistics of the dye
concentration was evaluated from space-time diagrams similar to
those in Fig.23. The regions near the walls of the channel with
the widths of $0.1d$ were excluded from the statistics, though,
because of possible image aberrations. In a stationary flow regime
($Wi$ below 3.2), when the concentration profile did not change in
time, the brightness profiles were measured over short time
intervals (about 100 s). In the regime of an irregular flow,
however, the profile of concentration was strongly fluctuating.
So, in order to obtain representative statistics of $c$, the
measurements of the brightness profile were taken during quite
long intervals of time (about 20-30 min), that typically
corresponded to the total liquid discharge of about $10^{3}d^{3}$.

The plot in Fig.24 is somewhat analogous to those in Fig.3 and in
Fig.18, which show dependence of the flow resistance on $Wi$.
Indeed, the decrease in $c_{rms}$ is an integral result of mass
transfer
 produced by the irregular flow in the channel, just as growth of the flow resistance is an
integral characteristic of increase of momentum transfer in the
elastic turbulence. At $Wi$ below 2 dye distribution patterns in
the polymer solution were similar to those in the solvent,
Fig.22a, with $c_{rms}$ close to unity. Some decrease in $c_{rms}$
at low $Wi$ is due to large residence time and diffusional
smearing of the boundary between the liquids.

At $Wi$ of about 2 reduction of $c_{rms}$ from 1 to about 0.9
occurred, which was a result of a transition in the flow. Although
this transition produced rather complex spatial distribution of
the dye (data not shown), the patterns were stationary and mixing
was rather minor. This transition was probably corresponding to
onset of stationary vortices with vorticity directed along the
mean flow, as it was predicted in Ref.\cite{Joo}. Those vortices
have recently been observed directly using particle image
velocimetry\cite{unpubl} (see also below). The most striking
feature of the plot in Fig.24 is certainly an abrupt drop in
$c_{rms}$ at $Wi_{c}=3.2$, where the irregular motion of the
liquid set in. It is worth noting here that since the mixing and
reduction in $c_{rms}$ is a cumulative effect of stirring and
diffusion in the flow starting from the inlet, the actual value of
$c_{rms}$ above the transition is defined by the distance from the
inlet. On the other hand, the onset of the fluctuating flow
occurred rather consistently along the whole channel, and we
expect $Wi_{c}$ to be virtually independent of the channel length,
when the latter is sufficiently large.

We studied dependence of $c_{rms}$ on $N$ at $Wi=6.7,$
corresponding to highest homogeneity of the mixture near the
channel exit (Fig.24), and found $c_{rms}$ to decay exponentially
with the distance from the inlet with a characteristic decay
length $\Delta N$ of about 15 segments\cite{mix}. One can learn
from Fig.24 that, if $Wi$ is raised above 6.7, $c_{rms}$ starts to
increase again. The most plausible explanation for this is
saturation of growth of the velocity fluctuations together with
reduction of the residence time in the flow at growing $Wi$ (and
average flow velocity, $\bar{V}=Q/d^2$). If the ratio between
fluctuating and average flow velocities remains constant, while
they both increase, the stirring in the flow remains the same, but
there is less time available for molecular diffusion, and
homogeneity is reduced as a result of it. This situation can be
qualitatively described by growth of the Peclet number,
$Pe=\bar{V}d/D$. It was recently found for a flow of a polymer
solution in a channel of the same shape and at similar $Wi$, that
the characteristic length $\Delta N$ increases as
$Pe^{0.25}$\cite{Teo}. This suggests that $c_{rms}$ should start
increasing with $Wi$ once growth of the velocity fluctuations is
slowed down.

Typical time of mixing in the channel at $Wi=6.7$ was found to be
3-4 orders of magnitude shorter than diffusion time, $d^{2}/D$,
for the small molecules of fluorescein\cite{mix}. Dependence of
the efficiency of mixing at the optimal flow conditions (for the
80 ppm solution it was $Wi=6.7$, Fig.24) on concentration of the
polymers was surprisingly weak (although $Wi_{c}$ grew fast, as
the polymer concentration was decreasing). So, for a solution with
the polymer concentration of 10 ppm ($\eta /\eta _{s}=1.03$),
$c_{rms}$  of as low as 0.29 could be reached at $N=29$. (It was
measured at $Re=0.065$, where inertial effects were still
negligible). Excitation of irregular flow and active mixing was
observed down to the polymer concentration of 7 ppm.

Fig.25 shows power spectra of fluctuations of longitudinal and transversal components of the velocity
in the polymer solution at $Wi=6.7$. The measurements were done at $N=12$, near the middle of the
half-ring in the middle of the channel. A spectrum of the velocity fluctuations in the flow of the
pure solvent at the same $Q$, giving just instrumental noise, is shown for comparison. The mean
velocity was $\bar{V}=6.6$ mm/s; the RMS of fluctuations, $V_{rms}$, was $0.09\bar{V}$ and
$0.04\bar{V}$ in the longitudinal and transversal directions, respectively. Measuring the transversal
velocity component in a few off-center positions, we found non-zero averages, which typically
persisted for a few minutes and changed their sign rather randomly in time. This situation can be
explained by the presence of persistent longitudinal vortical structures in the flow. These vortices
should be filling the whole channel cross-section, with their vorticity direction randomly jumping
between parallel and anti-parallel to the mean flow. Those vortices may first appear at $Wi=2$ but
remain stationary below $Wi_c=3.2$.

In order to measure resistance as a function of the flow rate, we
used a micro-fabricated channel with the same proportions but with
a 30 times smaller segment size\cite{micromixing}. It had $d=100$
$\mu $m, $R_{1}=100$ $\mu $m, $R_{2}=200$ $\mu $m and consisted of
46 segments. The polymer solution had the same concentration of 80
ppm by weight of the same PAAm sample. The Newtonian solvent was
significantly thinner, though, containing 35\% of sucrose and 1\%
NaCl, and having viscosity of $4.2\cdot 10^{-3}$ $Pa\cdot s$ at
room temperature of 22 $^\circ$C. The polymer solution viscosity
was $\eta =5.6\cdot 10^{-3}$ $Pa\cdot s$ at $\dot{\gamma}=50$
s$^{-1}$.

The flow was driven by a difference of hydrostatic pressures, $\Delta P$, applied between the inlet
and outlet. Dependence of the flow rate, $Q$, on $\Delta P$ was measured using an in-situ compensation
technique\cite{logic}(see details in Ref.\cite{micromixing}). Fig.26 shows resistance ratio $\Delta
P/\Delta P_{lam}$ as a function of $Q$, where $\Delta P_{lam}$\ is the pressure difference expected in
a laminar flow for a liquid with a viscosity, $\eta(\dot{\gamma})$, at the same $Q$. (Average
$\eta(\dot{\gamma})$ was calculated as $4Q/d^{3}$ again.) One can see a distinct transition and an
onset of non-linear growth of the flow resistance at $Q\approx 8.5$ nl/s, corresponding to
$\dot{\gamma}=34$ s$^{-1}$ and $Re\approx 0.017$. The low value of Re suggests that the transition is
of a purely elastic nature. Relaxation time, $\lambda $, of the polymer solution was low and could not
be measured directly. For a purpose of estimation we can assume $\lambda \sim \eta _{s}$ and plug into
this relation the values for the more viscous solution used in the table-top channel, $\lambda =1.4$ s
at $\eta _{s}=0.153$ $Pa\cdot s$. That gives $\lambda =0.04$ s and $Wi=1.3$ at the instability
threshold for the current solution, which is in reasonable agreement with the results in Fig.24. As
$Q$ increases to 60 nl/s (corresponding to $Wi\simeq 9$), the flow resistance growth reaches a factor
of 2.8, which is comparable with the CT flow, although significantly lower than in the flow between
two plates.

\bigskip

\subsection{ Discussion}

The experimental results in Fig.24 and Fig.26 show that above the
elastic instability threshold the flow of the polymer solution in
the channel exhibits two major features of turbulent flows: major
increase in the rate of mixing and in the flow resistance.
Further, more solid evidence for the turbulent character of the
flow is given by the power spectra of fluctuating velocity in
Fig.25. The spectra of the both longitudinal and transversal
velocity components do not exhibit any distinct peaks and have
broad regions of a power law decay with a power of about -3.3.
Since the power spectra in Fig.25 were measured in a point with a
high mean flow velocity (10 times higher than a characteristic
fluctuating velocity), we can use the Taylor hypothesis and argue
that the spectra in Fig.25 actually reflect spatial structure of
the flow. Then the power law decay region can be transferred to
the spatial domain, with the power of the velocity fluctuations
scaling as $P\sim k^{-3.3}$ with the wave number, $k$. The absence
of peaks in the spectra in Fig.25 and significant extension of the
frequency region, where $P$ follows the power law, are strong
evidence of fluid motion in a broad range of spatial scales and of
turbulent character of the flow. We also notice here that the
exponent of -3.3 in Fig.25 is very close to those measured in the
flow between two plates, Figs.5, 8, 16, which varied from -3.3 to
-3.6 depending on the position. So, one can suggest that decay of
the power of the velocity fluctuations with an exponent around
-3.5 is a rather general feature of the elasticity induced
turbulent flows. It does not appear in the radial velocity
spectrum of the CT flow (Fig. 19), though, possibly because of
abundance of the coherent axisymmetric structures (see Section
4B).

 The functional form of the velocity power spectra, $P\sim k^{-3.3}$,
deduced from Fig.25, suggests that the power of fluctuations of
velocity gradients scales as $k^{-1.3}$.  An integral of
$k^{-1.3}$ diverges at $k\rightarrow 0$ and converges at
$k\rightarrow \infty $. It means that the main contribution to the
fluctuations of the velocity gradients and the velocity
differences at all scales comes from the biggest eddies, having
dimensions of the whole system (diameter of the channel). This
conclusion has an immediate implication for mixing in the flow: it
should result in the same type of patterns and in functionally the
same statistics as in the case of a completely homogeneous flow,
$\vec{V}(\vec{r},t)=\vec{V}_{0}(t)+\frac{\partial V_{i}}{\partial
r_{j}}(t)\cdot (\vec{r}-\vec{r}_{0})$, randomly varying in time.

Such a flow is a realization of the so-called Batchelor regime of
mixing \cite{Bat}, and the problem of statistics of a tracer (dye)
distribution in it has been solved analytically
recently\cite{Shr,Fal,Fouxon}. The Batchelor regime occurs at
small scales (below the Kolmogorov dissipation
scale\cite{landau,Frish}) in the usual, high $Re$, turbulence, and
it is rather difficult to realize in laboratory otherwise.
Therefore, the elastic turbulent flow in the channel provided a
very convenient experimental system for quantitative study of
mixing in this regime \cite{mix}. The experimental results on
correlation functions and PDF of dye concentration, and on their
dependence on time of mixing agreed very well with the theoretical
predictions \cite{mix}. A practical message of the experiments is
that very viscous liquids can be efficiently mixed in curvilinear
channels at very low flow rates by adding high molecular weight
polymers at very low concentrations. This method of mixing, we
believe, can find some industrial and laboratory
applications\cite{patent}.

\bigskip

\section{ Concluding remarks.}

We studied flows of dilute solutions of a flexible high molecular
weight polymer in three different flow set-ups, which shared the
same feature of high curvature of the flow lines, quantitatively
expressed in large gap ratios, $d/R$. High viscosity of the
solvent and long polymer relaxation times ensured an elasticity
dominated flow regime, where effects of the non-linear
constitutive relation between polymer stress and rate of
deformation prevailed over the inertia related non-linearity of
the Navier-Stokes equation. As a result, all flow transitions were
induced by the non-linear elasticity, occurred at $Wi$ on the
order of unity and at very low $Re$. In all three systems, a flow
above the instability threshold was randomly fluctuating in time
and space, and exhibited some or all of the main features of
turbulence: fluid motion excited in a broad range of spatial and
temporal scales, and significant increase in the rates of momentum
and mass transfer (flow resistance and mixing). The extend of the
flow resistance growth varied rather substantially between the
systems (factors of 20, 4.2 and 2.8 for the flow between two
plates, the CT flow, and the flow in the curvilinear channel,
respectively) and so did the exponents in the power law decay
regions of the velocity spectra (being from -3.3 to -3.6 for the
flow between two plates, -1.1 and -2.2 for the CT flow, and -3.3
for the curvilinear channel). We believe, however, that all the
experimental findings are fitting rather well to the same general
framework of a phenomenon, which we call elastic turbulence.

Observation of the elastic turbulence and investigation of its
properties have been greatly facilitated by the choice of the
systems with large $d/R$, where the elastic instability occurs at
low $Wi$, and by the choice of the high molecular weight polymer,
which is only moderately extended (compared to its full contour
length) at the instability threshold. Nevertheless, we believe
that elements of the elastic turbulence should generally appear in
three-dimensional flows of polymer solutions at high $Wi$ and low
$Re$. Possible realizations of this regime may range from flows of
polymer melts in industrial reactors to small scale flows in drag
reducing aqueous polymer solutions. Dependence of the elastic
turbulence onset conditions on parameters of the system and on
properties of the polymers solution (discussed in Section III.D.2)
implies that using polymer solutions with sufficiently high
elasticity, one can excite turbulent motion at arbitrary low
velocities and in arbitrary small tanks. (In some cases it may be
the only way to produce a chaotic flow, which may be needed for
fluid mixing.) Irregularly fluctuating flows showing multiple
features of the elastic turbulence have been recently demonstrated
in a rotating flow between two plates with a 300 $\mu $m
gap\cite{corinne} and in a 100 $\mu$m thick curvilinear
channel\cite{micromixing}. An apparent restriction is, however,
that the size of the tank has still to be large compared to the
size of the polymer coils.

Recently there have been some serious advances in theory of random flows of polymer solutions in the
regimes of both high and low Re\cite{Volodya,Volodya2,lebedev,chertkov}. In particular, the following
explanation of the velocity spectra shapes in Figs.5, 8, 16, 25 was proposed\cite{Volodya2}. Fast
decay of the fluctuation power with $k$ implies a velocity field, where the main contribution to
deformation and stirring (stretching and folding) at all scales comes from randomly fluctuating
velocity field at the largest scale of the system. Therefore, it is suggested that the leading
mechanism for generation of small scale (high $k$) fluctuations in the elastic stress is advection of
the fluid (which carries the stress) in this fluctuating large scale velocity field. Hence, the
fluctuating velocity field and stress tensor can both be decomposed into large and small scale
components, and the leading mechanism for generation of the small scale (high $k$) portions is
advection by the fluctuating large scale flow. (The theory considers the elastic stress tensor being
passively advected in a random velocity field, that is analogous to the concept of a passively
advected vector in the magnetic dynamo theory \cite{Volodya2}.) The small scale velocity fluctuations
are a product of the small scale stress fields. The elastic stresses are mainly pumped by the large
scale flow field, and the smaller scale stress fields created by the advection are permanently
decaying because of the polymer relaxation. These simultaneously occurring advection and relaxational
decay of $\tau_p$ result in a quick decay of $\tau_p$ fluctuations at large $k$ that should produce
$P\sim k^{-\alpha}$ velocity spectra with $\alpha>3$. The current understanding of the elastic
turbulence is still remaining quite incomplete, however, and this phenomenon certainly awaits further
detailed experiments and numerical simulations.

\textbf{Acknowledgement.} We are grateful to M. Chertkov, G. Falkovich, and V. Lebedev for many useful
and illuminating discussions. This work is partially supported by an Israel Science Foundation grant,
Binational US-Israel Foundation, and by the Minerva Center for Nonlinear Physics of Complex Systems.


\subsection{List of notations.}

$c$ - passive tracer (dye) concentration,

$\bar{c}$ - average concentration,

$c_0$ - initial concentration of dye,

$c_{rms}$ - root mean square of deviations of dye concentration
from its average divided by the average,

$d$ - gap width,

$d/R$ - gap ratio,

$f$ - frequency,

$k$-wave number,

$K=(d/R)(\eta_p/\eta)Wi^2$ - parameter of elastic non-linearity
for normal stress driven instabilities in curvilinear flows,

$L$ - system size,

$M_w$ - average molecular weight,

$N$ - number of half-ring in a curvilinear channel,

$N_1$ - first normal stress difference,

$p$ - pressure,

$P$ - power of fluctuations,

$Pe$ - Peclet number,

$Q$ - flow discharge rate,

$r,\phi,z$ - cylindrical coordinates (radial, azimuthal, axial),

$R$ - radius of either a plate in rotating flow between two plates
or inner cylinder in Couette-Taylor flow,

$R_g$ - radius of gyration of a polymer,

$Re=VL/\nu$-Reynolds number,

$t_{vd}=d^2/\nu$ - viscous diffusion time,

$T$ - torque,

$T_{lam}$ - torque in polymer solution flow in a laminar regime,

$T_s$ - torque measured in a solvent flow,

$Ta=(d/R)Re^2$ - Taylor number,

$V_r$ - radial (spanwise) velocity component, $D$-diffusion
coefficient,

$V_{\phi}$ - azimuthal (longitudinal) velocity component,

$Wi$ - Weissenberg number,

$Wi_c$ - critical $Wi$ of the elastic instability onset,

$z_b$ - characteristic thickness of boundary layer,

$\dot{\gamma}$ - shear rate,

$\lambda$ - polymer relaxation time,

$\eta$ - viscosity,

$\eta_s$ - viscosity of Newtonian solvent,

$\eta_p$ - part of solution viscosity due to polymer molecules,

$\eta'$ - in-phase component of viscosity in an oscillatory test,

$\eta''$ - out-of-phase component of viscosity in an oscillatory
test,

$\nu$ - kinematic viscosity,

$\rho$ - density,

$\bf{\tau}$ - stress tensor,

$\bf{\tau_s}$ - stress tensor due to Newtonian solvent,

$\bf{\tau_p}$ - polymer stress tensor,

$\tau_w$ - shear stress near a wall,

$\tau_w^{lam}$ - shear stress near a wall in a laminar flow,

$\Omega$ - rotation (angular) velocity.



\begin{figure}

\caption{Schematic drawing of the experimental set-up. The set-up
consists of a stationary cylindrical cup with a flat bottom (the
lower plate), which is concentric with the rotating upper plate. A
special cover is put from above to minimize evaporation of the
liquid.}

\label{figa}
\end{figure}

\begin{figure}

\caption{Upper curve: dependence of the apparent viscosity,
$\eta(\dot{\gamma})$, of 80 ppm PAAm solution in 65\% sucrose, 1\%
NaCl in water at 12 $^\circ$C on shear rate (semi-logarithmic
coordinates). Applied shear rate was gradually increased during
the test. Lower curve shows a similar run for the pure solvent.
{\it Inset:} apparent relaxation time, $\lambda$, as a function of
angular frequency, $\omega$, measured for shear rate oscillations
with an amplitude of 1 s$^{-1}$.}

\label{figa2}
\end{figure}

\begin{figure}

\caption{The ratio of the average stress at the upper plate, $\tau_w$, measured in the flow, to the
stress, $\tau_w^{lam}$, in imaginary laminar shear flow
with the same boundary conditions, as a function of the shear rate, $\dot{%
\gamma}$. The curves 1 and 2 are for the polymer solution flow with the gap $%
d=10$ mm and $20$ mm, respectively. The shear rate was gradually
varied in time. Thin black lines represent increasing
$\dot{\gamma}$; thick gray lines represent decreasing
$\dot{\gamma}$. Curve 3 is for the pure solvent.}

\label{figb}
\end{figure}

\begin{figure}

\caption{ Power spectra of fluctuations of the angular velocity,
$\Omega$, of the upper plate at different applied torques, $T$.
Curves 1 - 5 correspond to average shear rates of 1.25, 1.85, 2.7,
4 and 5.9 s$^{-1}$, respectively (all above the transition point
$\dot{\gamma}\simeq1$ s$^{-1}$, Fig.3). The
power, P, of fluctuations is fitted by a power law, $P\sim f^{-4.3}$, for $%
\dot{\gamma}=4$ s$^{-1}$ over about a decade in frequencies, $f$.
Curve 6, taken with the pure solvent at a shear rate of 4
s$^{-1}$, shows instrumental noise. }

\label{figc}
\end{figure}

\begin{figure}

\caption{ Power spectra of velocity fluctuations in the standard set-up at different shear rates,
$\dot{\gamma}$. The fluid velocity was measured by
LDV in the center of the flow. The curves 1 - 5 correspond to $\dot{\gamma}$%
=1.25, 1.85, 2.7, 4, and 5.9 $s^{-1}$, respectively (all above the transition point
$\dot{\gamma}\simeq 1$, Fig.3). The power, P, of fluctuations is fitted by a power law, $P \sim
f^{-3.5}$, for $\dot{\gamma}=4 s^{-1}$ over about a decade in frequencies, $f$.}

\label{figd}
\end{figure}

\begin{figure}

\caption{Representative snapshots of the flow taken from below.
Field of view corresponds to the upper plate area. The flow was
visualized by seeding the fluid with light reflecting flakes.
(a-b) the polymer solution at $Wi=6.5$, $Re=0.35$; (c-e) the polymer solution at $Wi=13$, $%
Re=0.7$; (f) the pure solvent at $Re=1$.}

\label{fige}
\end{figure}

\begin{figure}

\caption{Average Fourier spectra of the brightness profiles taken along the diameter (thin black line)
and along the circumference at a radius of $2d$ (thick gray line). The averaging was made over about
20 minutes of the flow time and about 2000 snapshots.}

\label{figf}
\end{figure}

\begin{figure}

\caption{ Power spectra of fluctuations of radial velocity, $V_r$, at $\dot{%
\gamma}=4$ $s^{-1}$ measured at $z/d=0.5$ at different radii.
Curves 1 - 4 correspond to $r=0$, $r=d/2$, $r=2d$ and $r=3d$,
respectively. The average flow velocities, $(V_{\phi}, V_r)$, in
mm/s were (0,0), (0.13, 0.19), (3.81, 1.17), (6.99, 0.89) for the
curves 1 - 4, respectively.}

\label{figg}
\end{figure}

\begin{figure}

\caption{Snapshots of consecutive stages of mixing of a droplet of ink in the polymer solution in the
half-size set-up, view from below. The are of the photographs corresponds to the area of the white
upper plate. Rotation of the upper plate at $\Omega=1.47$ s$^{-1}$ ($\dot{\gamma}=5.6$ s$^{-1}$) was
suddenly started at $t=0$.}

\label{figh}
\end{figure}

\begin{figure}

\caption{ Snapshots of consecutive stages of mixing of a droplet of ink in the pure solvent in the
half-size set-up, view from below. Rotation of the upper plate at $\Omega=1.47$ s$^{-1}$ was suddenly
started at $t=0$.}

\label{figi}
\end{figure}

\begin{figure}

\caption{ Average azimuthal velocity, $\bar{V}$, $y$-axis on the left, curves
1-3, and RMS of fluctuations of the azimuthal velocity, $V_{rms}$, $y$%
-axis on the right, curve 4, as functions of the distance, $z$,
from the upper plate. The measurements were done at $r=2d$. The
average velocities are divided by the upper plate velocity at
$r=2d$. Curve 1 - polymer solution at $\dot{\gamma}=2.7$ s$^{-1}$;
curves 2,4 polymer solution at $\dot{\gamma}=4$ s$^{-1}$ (see
Fig.3-5, 8); curve 3 - pure solvent at $\dot{\gamma}=4$ s$^{-1}$,
$Re\simeq 1.2$. The RMS of velocity fluctuations in the polymer
solution, curve 4, is multiplied by $\lambda/d$ to make it
dimensionless.}

\label{figj}
\end{figure}

\begin{figure}

\caption{Probability distribution functions (PDF) of flow velocity measured at $r=2d$, $z=d/4$ and
$\dot{\gamma}=4$ s$^{-1}$, circles. Solid lines
represent fits by Gaussians with some skewness. (a) azimuthal velocity, $%
V_{\phi}$; (b) radial velocity, $V_r$.}

\label{figk}
\end{figure}

\begin{figure}

\caption{Probability distribution functions (PDF) of velocity gradients measured at $r=2d$, $z=d/4$
and $\dot{\gamma}=4$ s$^{-1}$, circles. The velocity gradients are made dimensionless by
multiplication by the relaxation time, $\lambda $. Solid lines represent Gaussian fits. (a)
longitudinal gradient, ${\frac{\partial V_{\phi }}{r\partial \phi }}$; (b) transversal gradient
${\frac{\partial V_{r}}{r\partial \phi }}$.}

\label{figl}
\end{figure}
\begin{figure}

\caption{Dependence of the upper plate stress ratio, $\tau
_{w}/\tau _{w}^{lam} $, (actual polymer solution flow and a
laminar flow with the same boundary conditions, cf. Fig.3) on the
shear rate for a set-up with dimensions reduced four-fold compared
with the standard configuration of the flow between two plates.
Black and gray curves correspond to raising and decreasing shear
rate respectively.}

\label{figm}
\end{figure}

\begin{figure}

\caption{Mixing patterns in polymer solution flow in set-ups with flow between two plates and with
dimensions reduced two-fold (left column) and four-fold (right column) compared with the standard
configuration. Photographs in consecutive rows were taken at equal times elapsed from sudden inception
of the upper plate rotation at $\Omega=1.47$ s$^{-1}$ ($\dot{\gamma}=5.6$ s$^{-1}$). Images in the
right column are magnified by a factor of two to match the sizes. The area of the photographs
corresponds to the area of the white upper plate. }

\label{fign}
\end{figure}

\begin{figure}

\caption{Power, $P$, of velocity fluctuations in set-ups with flow
between two plates as function of frequency, $f$. Curves 1 and 3
are for the radial velocity component, $V_{r}$, and curves 2 and 4
are for the azimuthal velocity, $V_{\phi}$. Black curves (1 and 2)
are for the standard set-up, and gray curves (3 and 4) are for a
set-up with two-fold reduced dimensions.}

\label{figo}
\end{figure}

\begin{figure}

\caption{Schematic drawing on the Couette-Taylor set-up, mounted
on top of the AR-1000 rheometer. A circular cap with a hole for
the rotating shaft, which covered the glass cylinder from above,
is not shown.}

\label{figp}
\end{figure}

\begin{figure}

\caption{The ratio of average shear stress in the actual flow of the polymer
solution and the stress corresponding to a laminar flow with the same $\dot{%
\gamma}$, $\tau /\tau _{lam}$, as a function of the Weissenberg number, $%
Wi=\lambda R_{1}\Omega /d$. Thin black line: increasing rotation rate, starting from $Wi=0.7$; thick
gray line: decreasing rotation rate, starting from $Wi=7$.}

\label{figq}
\end{figure}

\begin{figure}

\caption{Power of fluctuations of the radial component of the flow velocity, $%
V_{r}$, measured at $r=(R_{1}+R_{2})/2$, as a function of
frequency, $f$. Curves 1-4 correspond to $Wi=$ 8.5, 13.6, 21.8 and
35, respectively. Two power law fits for curve 4 in different
regions are shown to guide the eye.}

\label{figr}
\end{figure}

\begin{figure}

\caption{Probability distribution functions of the radial component of the flow velocity, $V_{r}$,
measured at $r=(R_{1}+R_{2})/2$, at different $Wi$. Curves 1-4 correspond to $Wi=$ 5.4, 13.6, 21.8 and
35, respectively.}

\label{figs}
\end{figure}

\begin{figure}
\caption{Schematic drawing of the curvilinear channel showing the inlet, a region of observation, and
the outlet.}

\label{figt}
\end{figure}

\begin{figure}
\caption{Photographs of the flow taken with the laser sheet
visualization (Fig.21) at different $N$. The field of view is 3.07
by 2.06 mm, and corresponds to the region shown in Fig.21 (rotated
$90^\circ$ counterclockwise). Bright regions correspond to high
concentration of the fluorescent dye. ($\bf a$) flow of the pure
solvent at $N=29$; ($\bf b, c, d$) flow of the polymer solution at
$Wi=6.7$ and at $N=8, 29, 54$, respectively.}

\label{figru}
\end{figure}

\begin{figure}

\caption{Representative space-time diagrams of the polymer solution flow at $Wi=6.7$ taken at
different positions, $N$, along the channel.}

\label{figv}
\end{figure}

\begin{figure}
\caption{Dependence of $c_{rms}$ (normalized root mean square of concentration deviations from the
average) on the Weissenberg number, $Wi$, measured near the channel exit at $N=29$ (semi-logarithmic
coordinates).}

\label{figw}
\end{figure}

\begin{figure}
\caption{Power, $P$, of fluctuations of velocity in the middle of the channel at $N=12$ as a function
of frequency, $f$.  The spectra in the polymer solution flow at $Wi=6.7$ for the velocity components
along and across the mean flow are shown by curves 1 and 2, respectively.  Curve 3 shows velocity
spectrum across the mean flow for the pure solvent at the same $Q$.}

\label{figx}
\end{figure}

\begin{figure}
\caption{ Ratio of a pressure drop across the channel in the
polymer solution flow, $\Delta P$, to a pressure drop for a
laminar flow, $\Delta P_{lam}$, as a function of rate of liquid
discharge, $Q$ (in semi-logarithmic coordinates).}

\label{figy}
\end{figure}
\end{multicols}

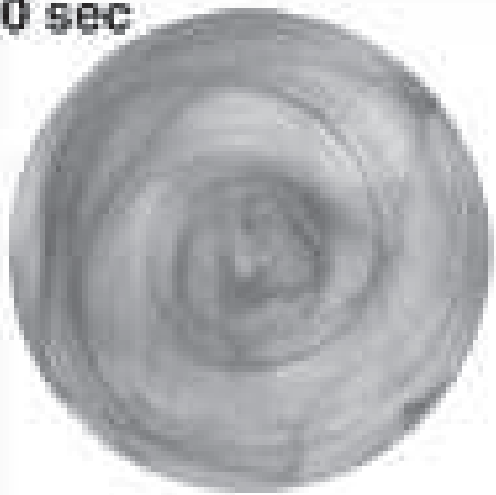
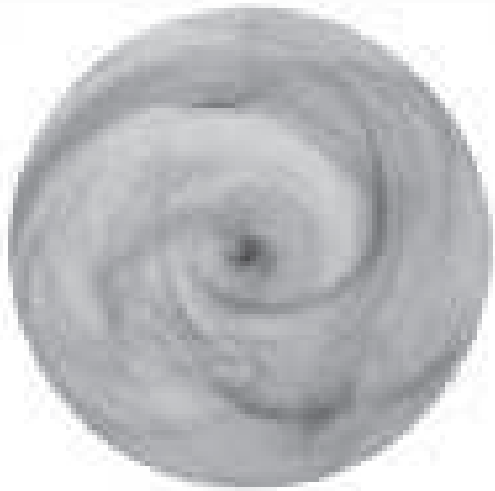
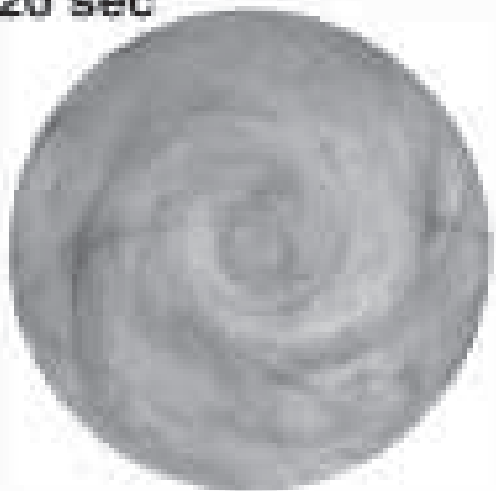
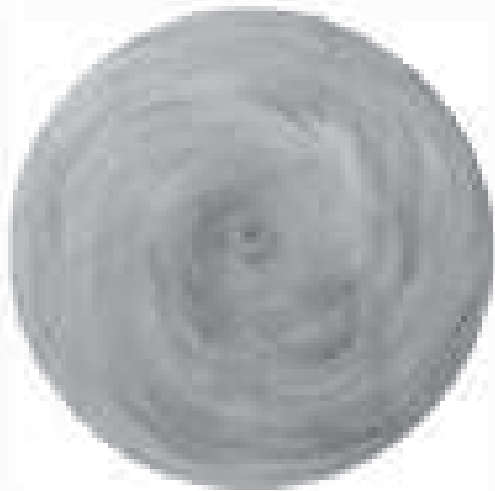
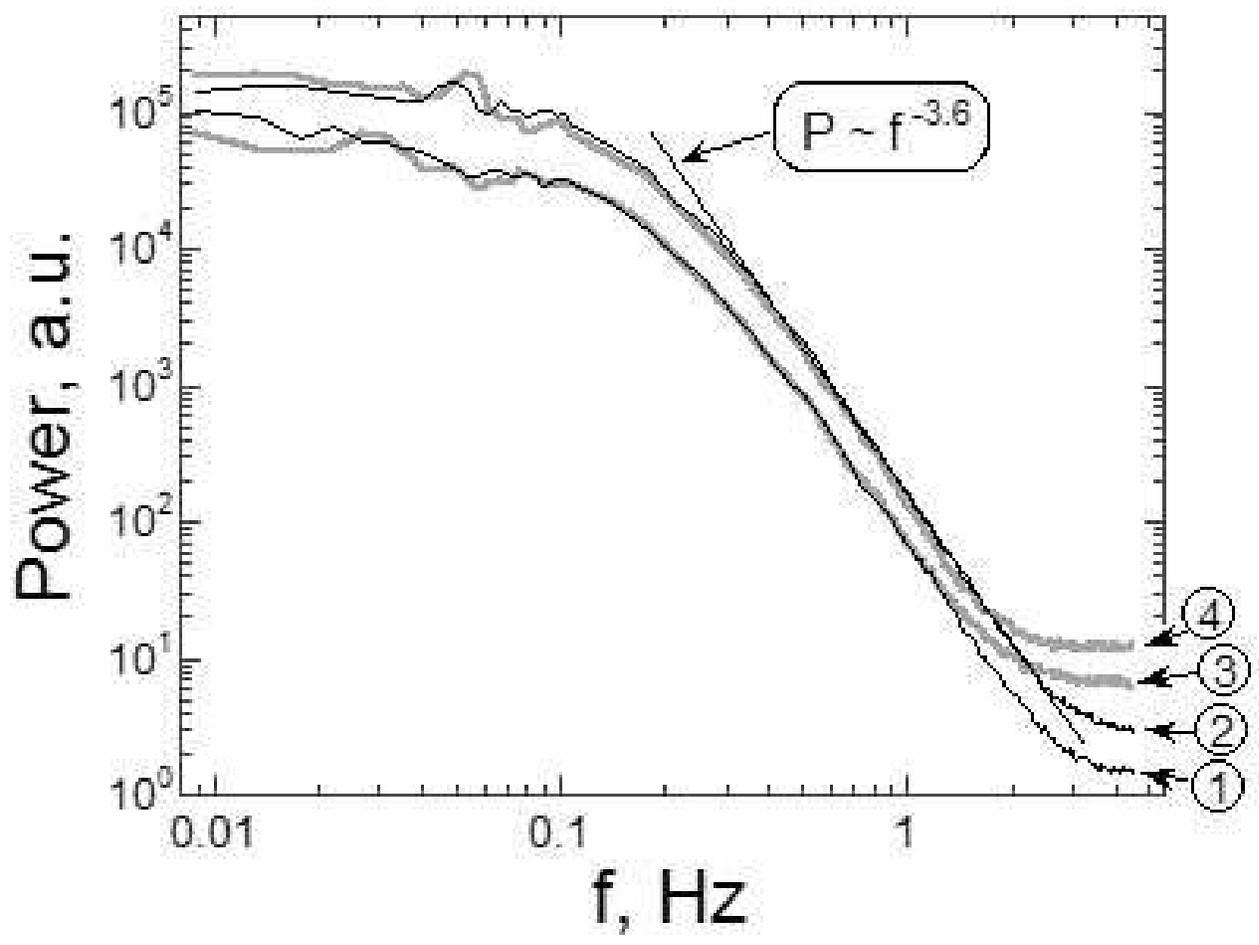
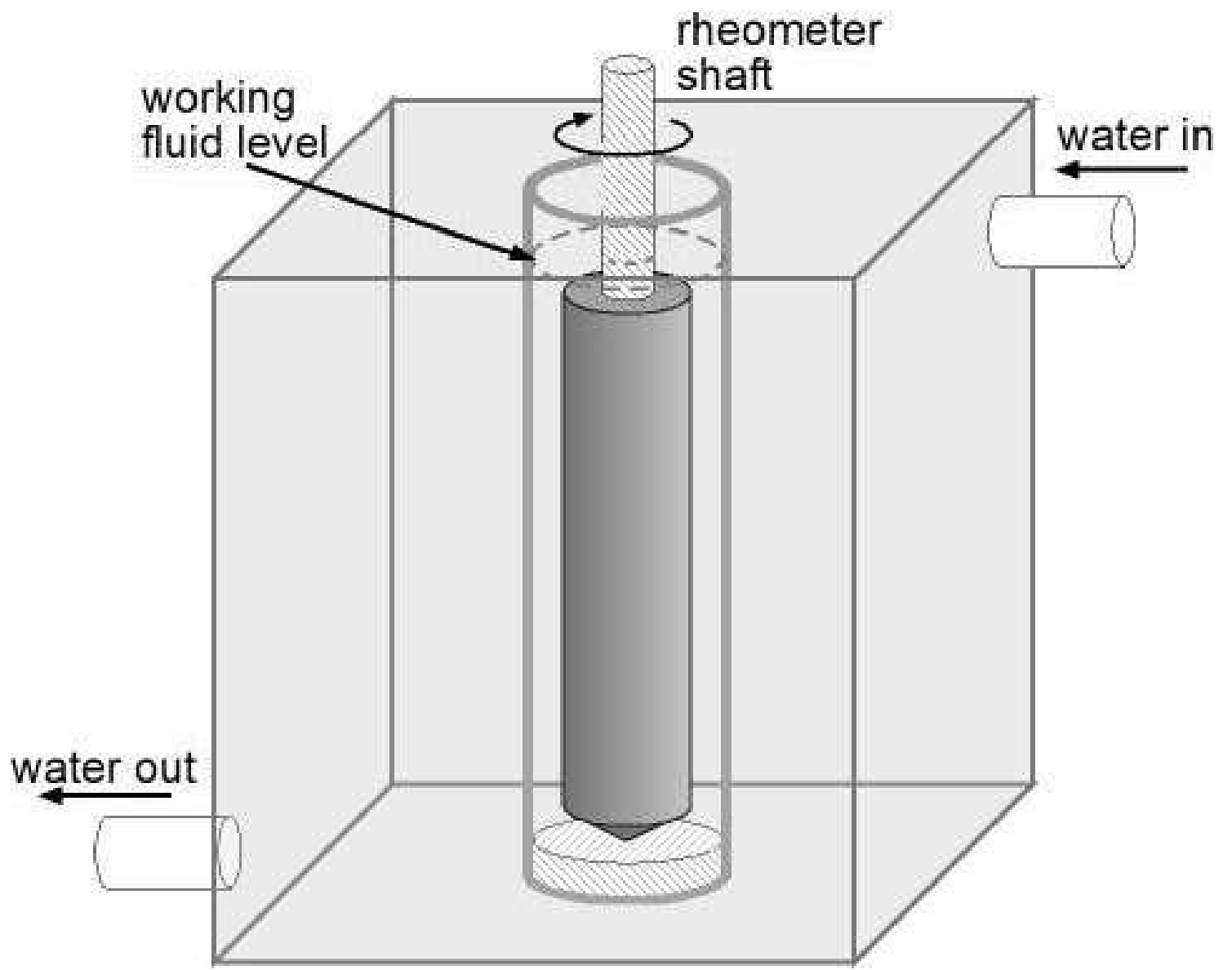
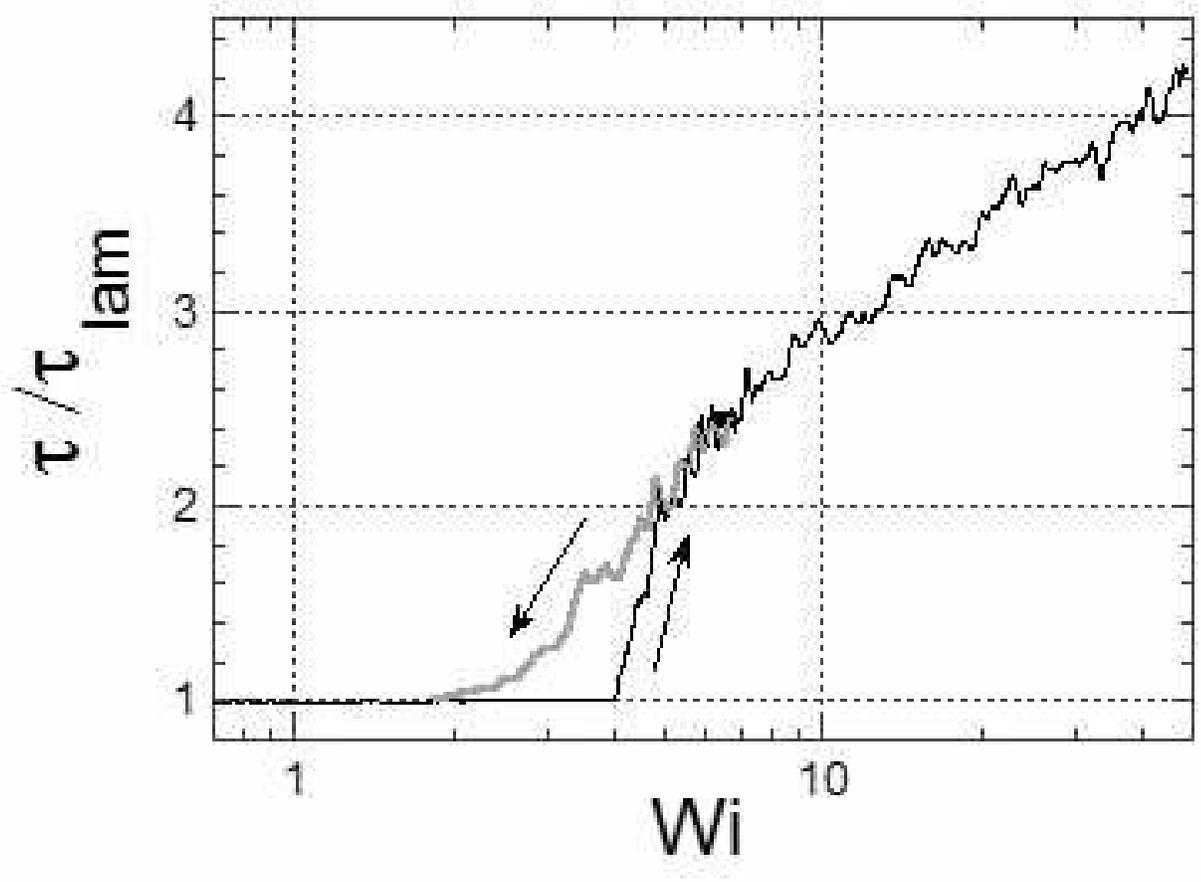
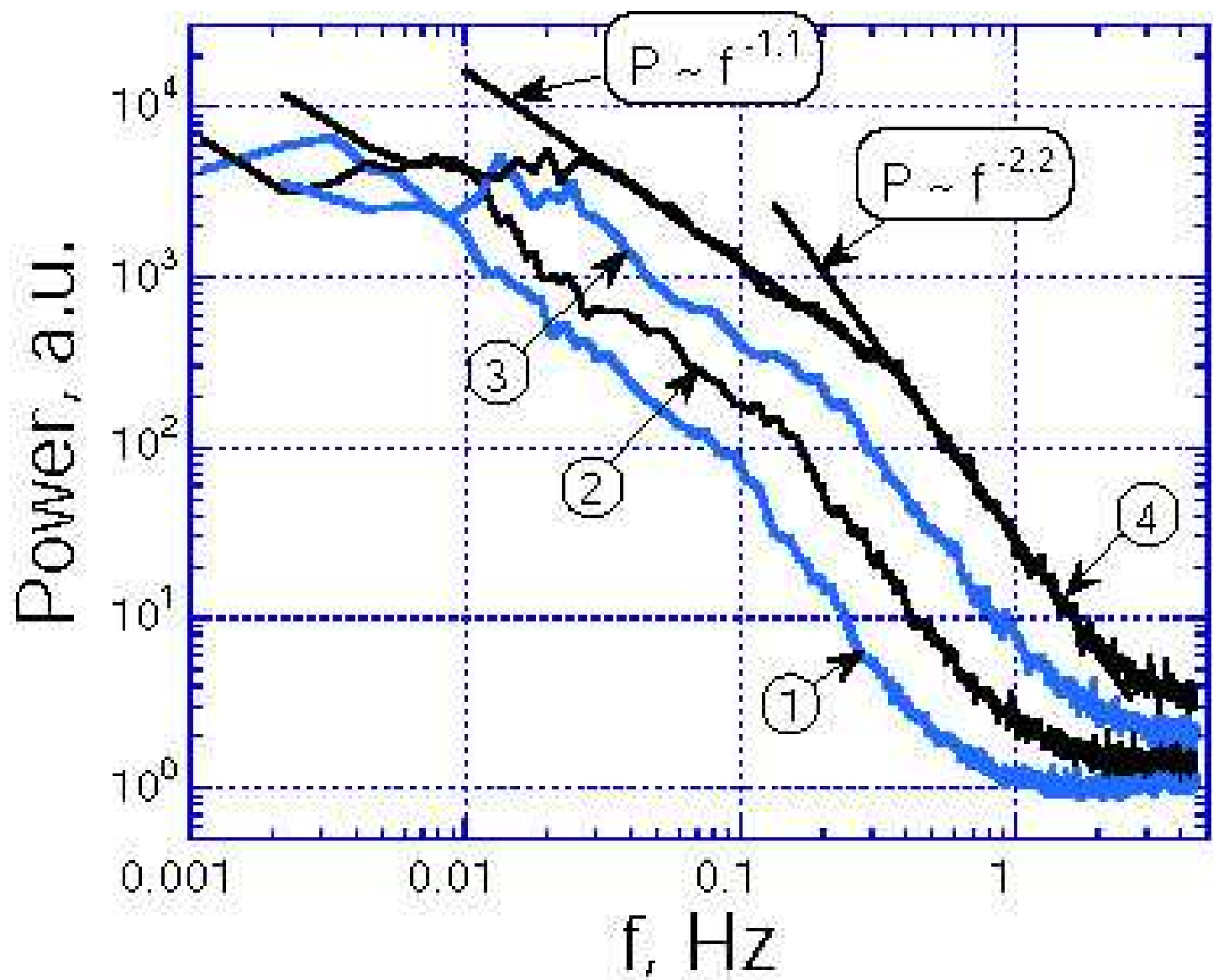
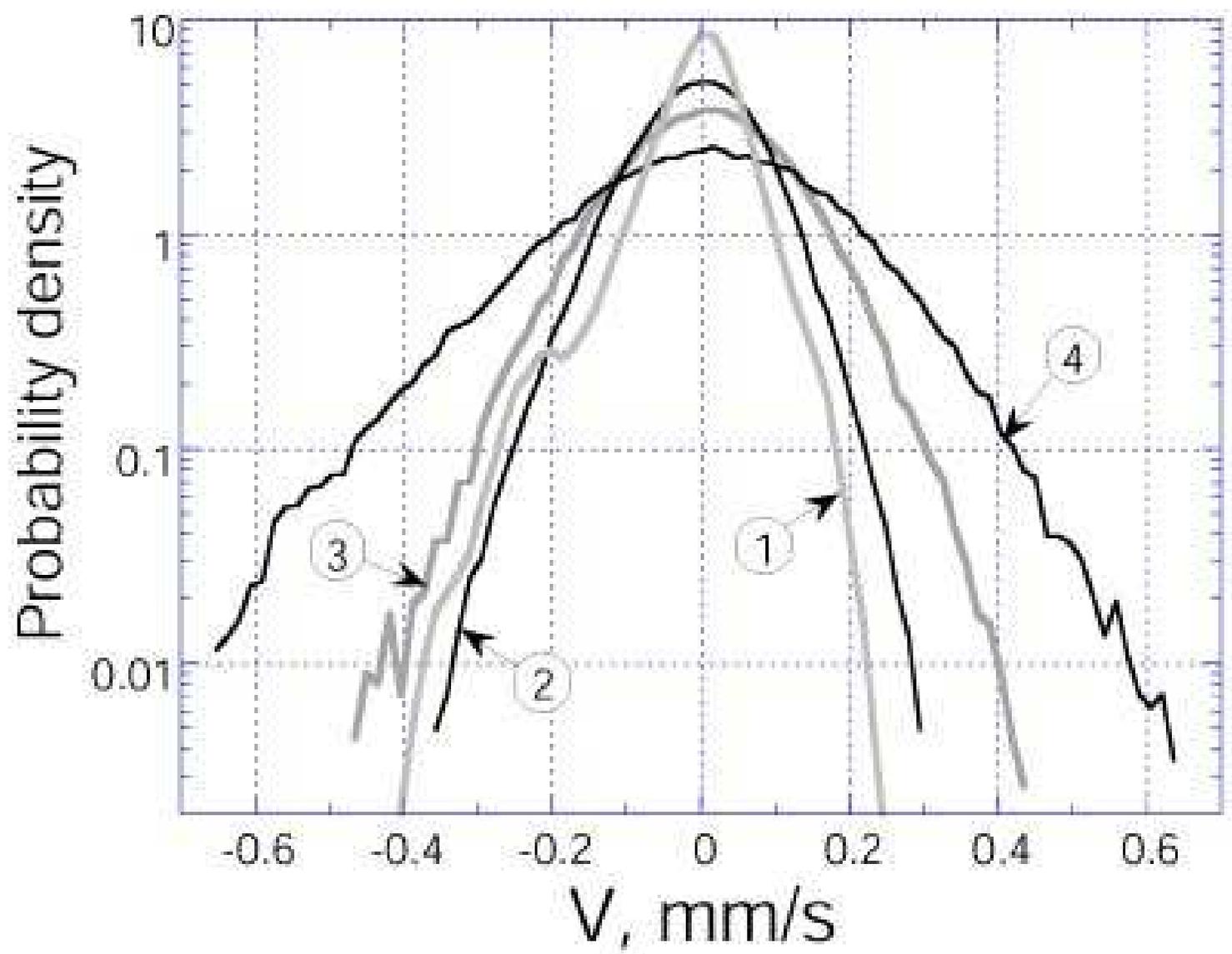
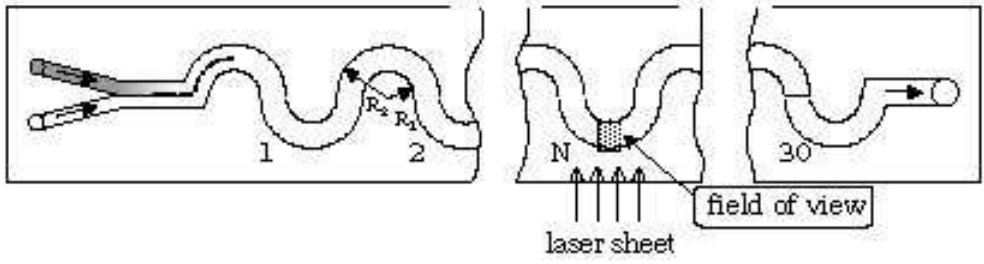
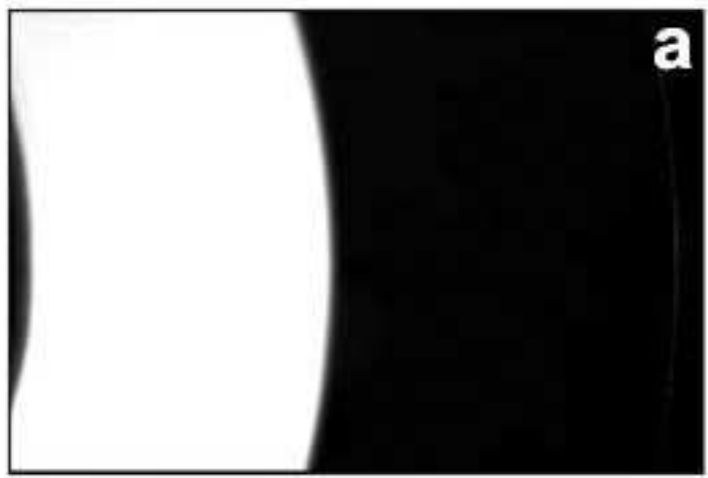
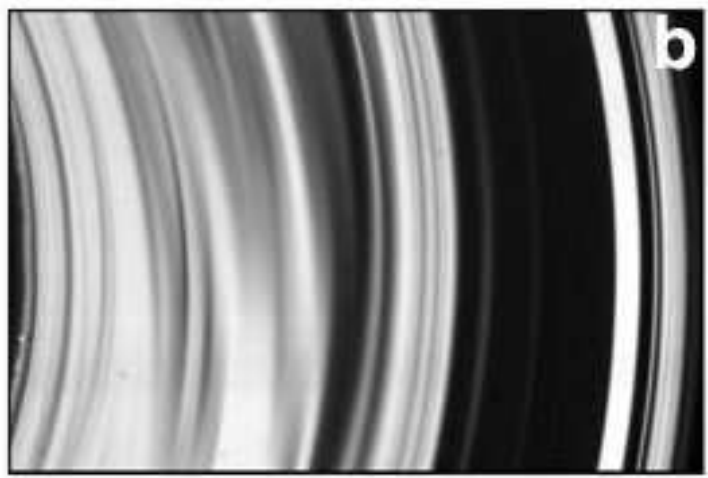
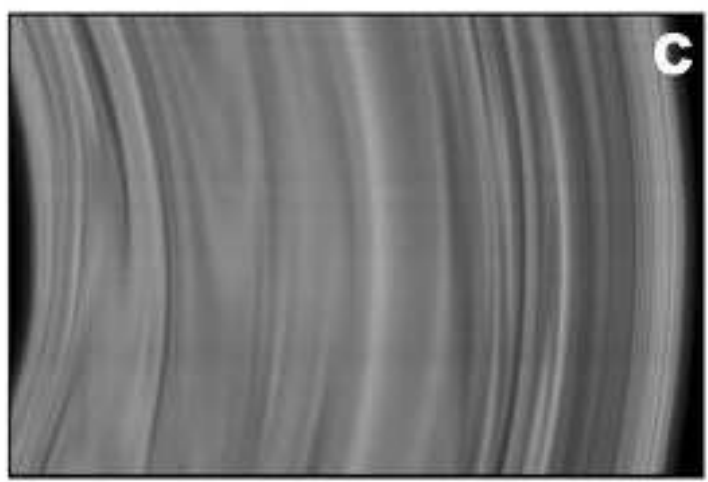
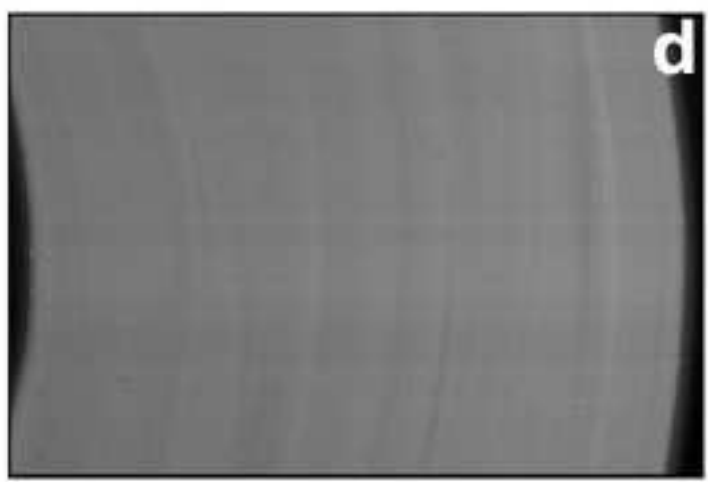
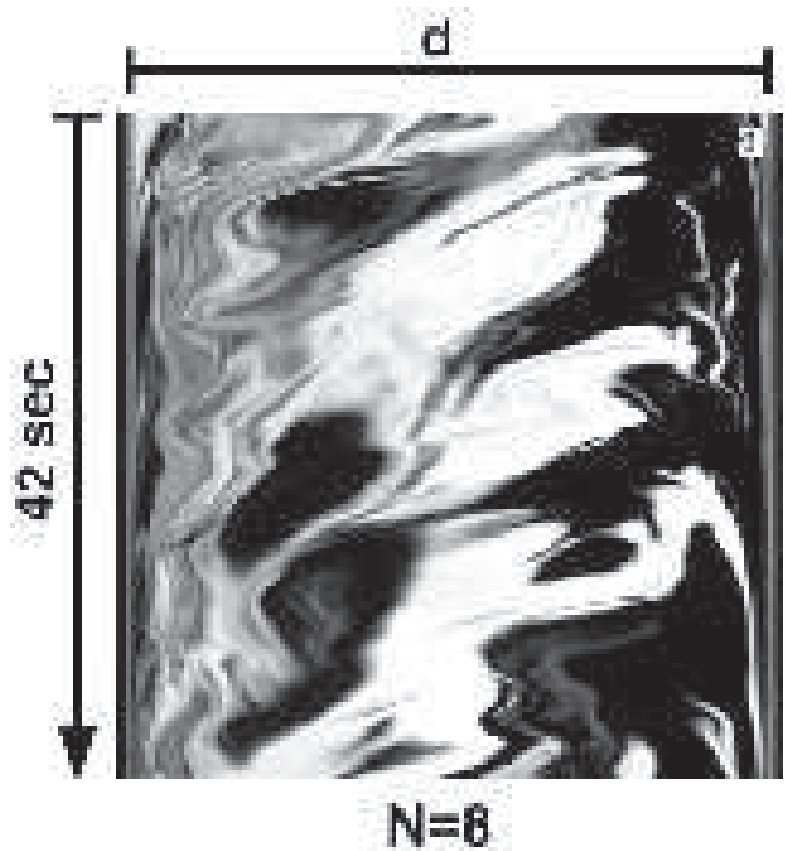
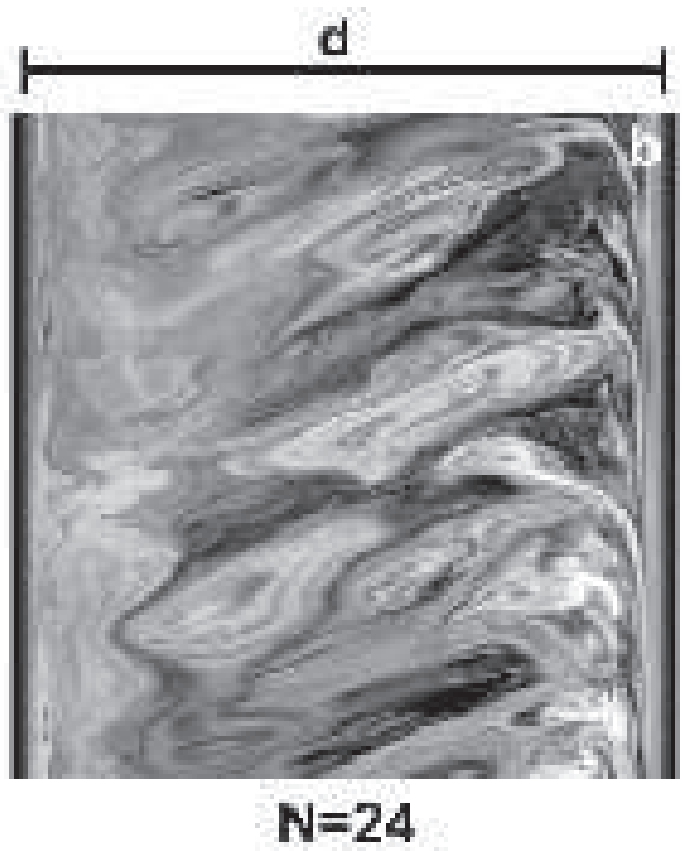
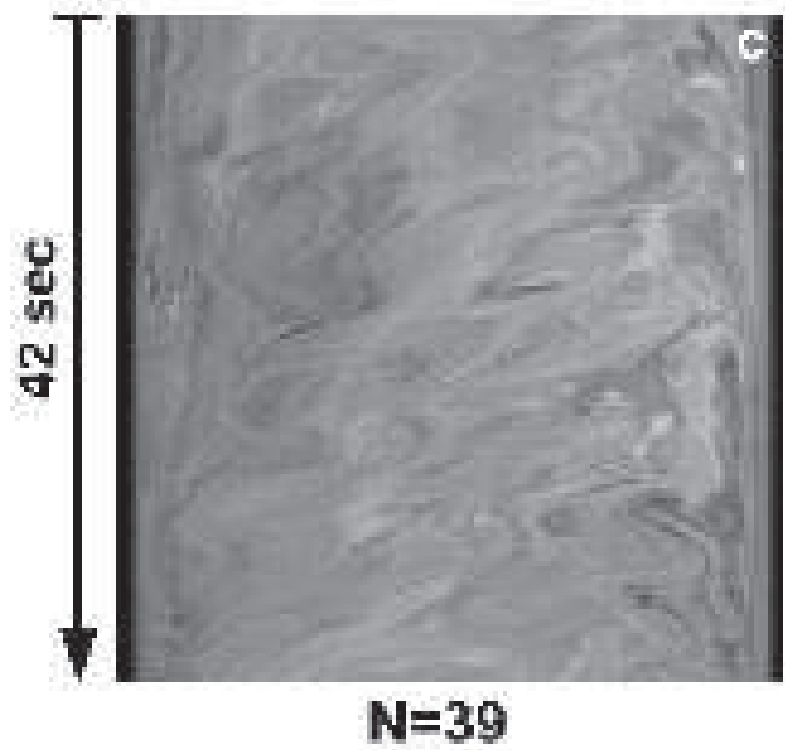
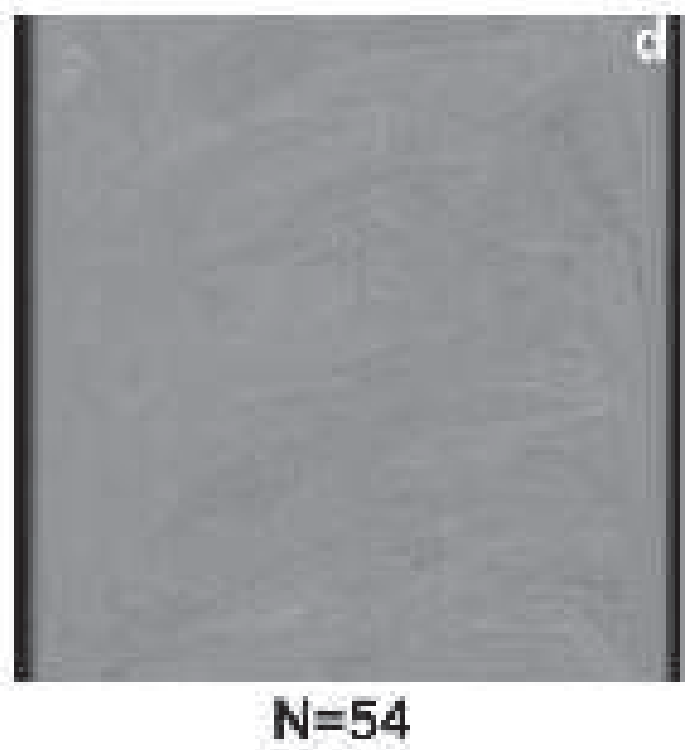
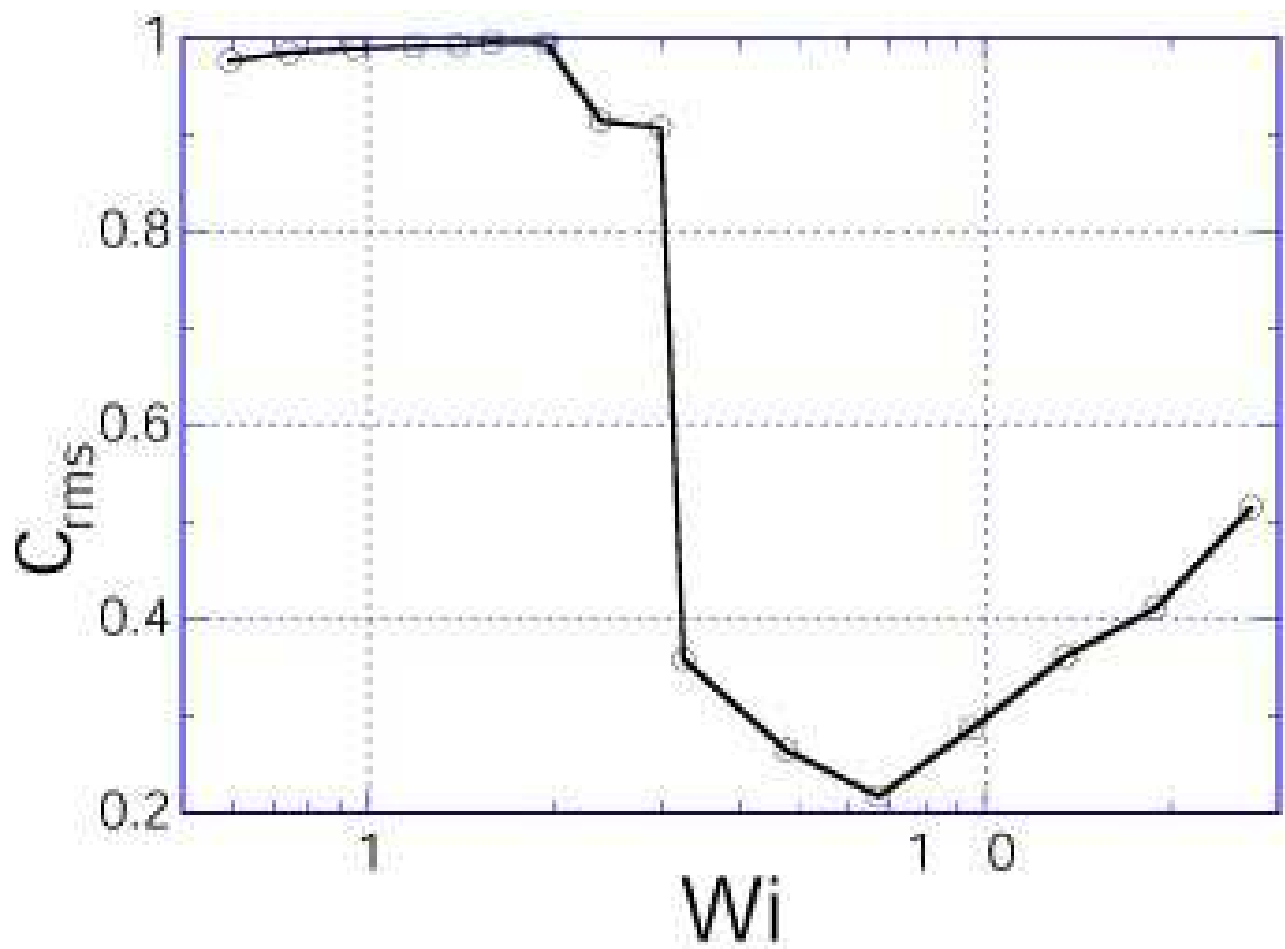
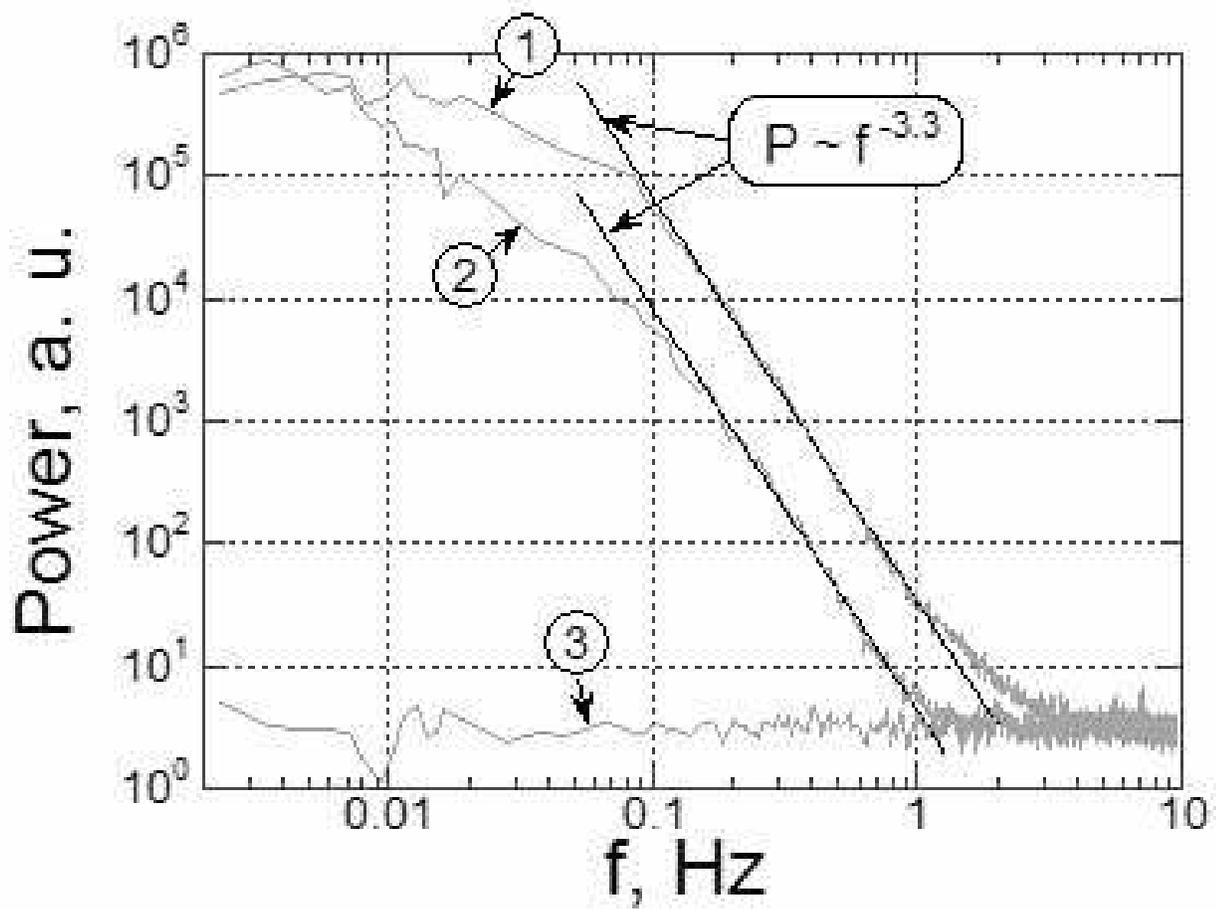
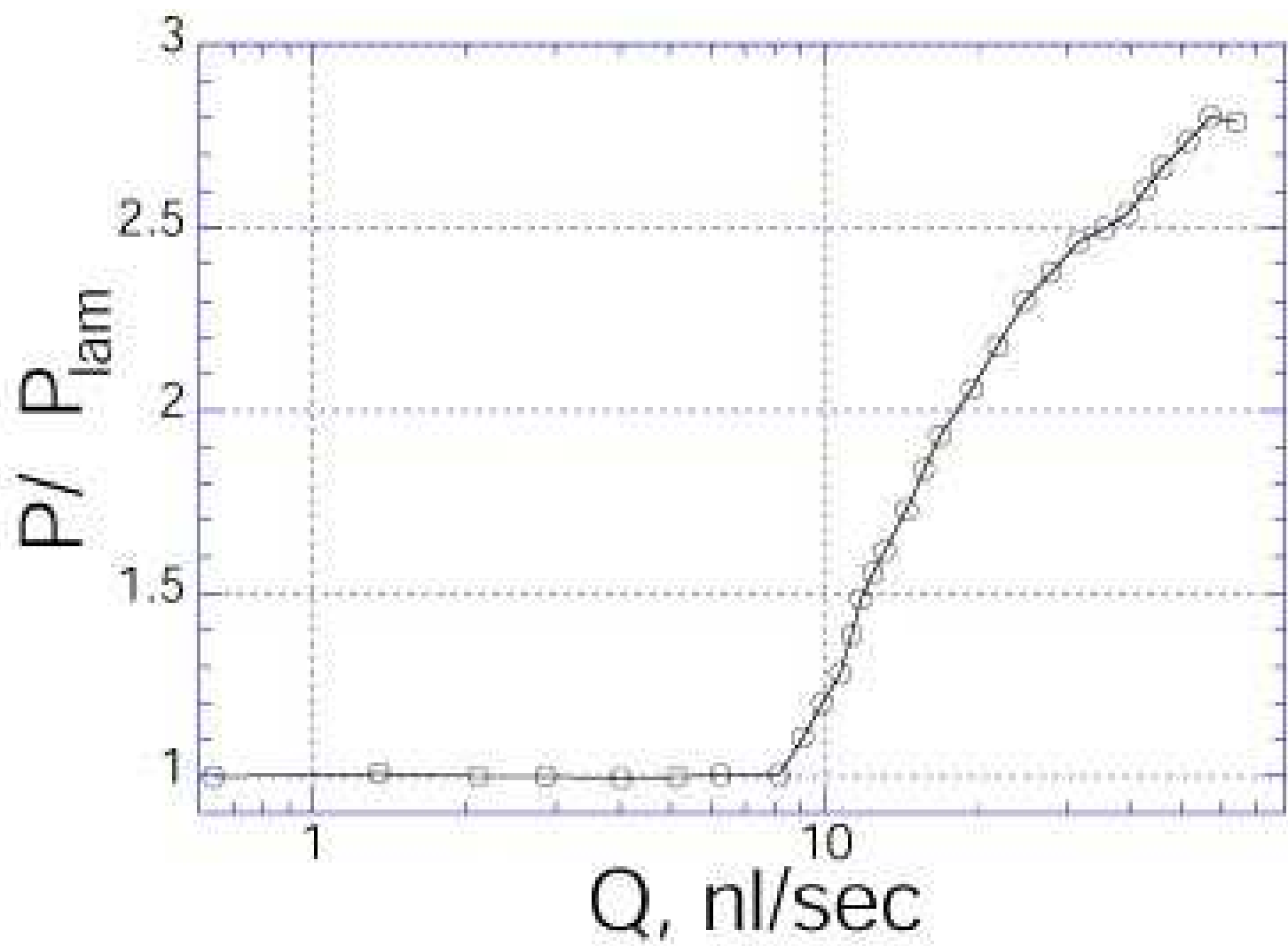
\end{document}